\documentclass[12pt]{iopart}
\usepackage{epsf}

\newcommand{\pd}{{\phantom{\dagger}}}

\begin{document}

\title[Quantum phase transitions]{Quantum phase transitions}

\author{Matthias Vojta
\footnote[3]{
To whom correspondence should be addressed
(vojta@tkm.physik.uni-karlsruhe.de)}
}

\address{Institut f\"ur Theorie der Kondensierten Materie,
Universit\"at Karlsruhe, \\
Postfach 6980, D-76128 Karlsruhe, Germany}

\begin{abstract}
In recent years, quantum phase transitions have attracted the interest
of both theorists and experimentalists in condensed matter physics.
These transitions, which are accessed at zero temperature by variation
of a non-thermal control parameter, can influence the behavior
of electronic systems over a wide range of the phase diagram.
Quantum phase transitions occur as a result of competing ground
state phases. The cuprate superconductors which can be tuned from a
Mott insulating to a $d$-wave superconducting phase by carrier doping
are a paradigmatic example.
This review introduces important concepts of phase transitions and
discusses the interplay of quantum and classical fluctuations near
criticality.
The main part of the article is devoted to bulk quantum phase transitions
in condensed matter systems.
Several classes of transitions will be briefly reviewed, pointing
out, e.g., conceptual differences between ordering transitions in
metallic and insulating systems.
An interesting separate class of transitions are boundary phase
transitions where only degrees of freedom of a subsystem become
critical; this will be illustrated in a few examples.
The article is aimed on bridging the gap between high-level theoretical
presentations and research papers specialized in certain classes of
materials. It will give an overview on a variety of different quantum
transitions, critically discuss open theoretical questions, and
frequently make contact with recent experiments in condensed matter
physics.
\end{abstract}



\maketitle

\tableofcontents

\newpage


\markboth{\uppercase{Quantum Phase Transitions}}{Quantum Phase Transitions}

\section{Introduction}

Phase transitions play an essential role in nature.
Everyday examples include the
boiling of water or the melting of ice, more complicated is
the transition of a metal into the superconducting state
upon lowering the temperature.
The universe itself is thought to have passed through several phase
transitions as the high-temperature plasma formed by the Big Bang cooled
to form the world as we know it today.

Phase transitions occur upon variation of an external control parameter;
their common characteristics is a qualitative change in the system
properties.
The phase transitions mentioned so far occur at finite temperature;
here macroscopic order (e.g. the crystal structure in the case of melting)
is destroyed by thermal fluctuations.
During recent years, a different class of phase transitions has
attracted the attention of physicists, namely transitions taking place
at zero temperature.
A non-thermal control parameter such as pressure, magnetic field,
or chemical composition, is varied to access the transition point.
There, order is destroyed solely by quantum fluctuations
which are rooted in the Heisenberg uncertainty principle.

Quantum phase transitions \cite{sondhi,bkreview,book,laughlin} have become a topic of
vivid interest in current condensed matter physics.
At first glance it might appear that the study of such special points
in the phase diagram is a marginal problem of interest only to specialists,
as such transitions occur at only one special value of a control
parameter at the experimentally impossible temperature of absolute zero.
However, experimental and theoretical developments in the last decades have
clearly established the contrary.
They have made clear that the presence of such
zero-temperature quantum critical points holds the key to so-far unsolved puzzles
in many condensed matter systems.
Examples include rare-earth magnetic insulators \cite{BRA96},
heavy-fermion compounds \cite{HeavyF,Lohn96}
high-temperature superconductors \cite{hightc,sciencereview},
and two-dimensional electron gases \cite{sondhi,KMBFPD95}.

As we will see below, quantum critical behavior, arising from the peculiar excitation
spectrum of the quantum critical ground state,
can influence measurable quantities over a wide range of the phase diagram.
The physical properties of the quantum fluctuations, which can
destroy long-range order at absolute zero,
are quite distinct from those of the thermal fluctuations responsible for
traditional, finite-temperature phase transitions.
In particular, the quantum system is described by a complex-valued wavefunction,
and the dynamics of its phase near the quantum critical point requires novel theories
which have no analogue in the traditional framework of phase transitions.

This article is intended as primarily non-technical introduction into
the field of quantum phase transitions, and could serve as a reference
for both theorists and experimentalists interested in the field.
In Sec.~\ref{sec:qpt} we start with summarizing the basics of finite-temperature
phase transitions, extend those general concepts to $T=0$ highlighting the
interplay between classical and quantum fluctuations near a quantum critical
point, and illustrate the correspondence between quantum transitions in
$d$ and classical transitions in $D=d+z$ dimensions, where $z$ is the dynamic
critical exponent.
In addition, we briefly discuss some aspects of the theoretical description of
phase transitions, like order parameter field theories and their renormalization
group analysis.
Sec.~\ref{sec:bulk} is devoted to the concrete description of several classes
of bulk quantum phase transitions, where ``bulk'' refers to the fact that the
whole system becomes critical.
The discussion includes conventional transitions involving spin or charge
order as well as metal--insulator and superconductor--insulator
transitions; it is intended to summarize some of the settled and open issues in
the field, and to make contact with experiments.
For more technical details the reader is referred to the references.
In Sec.~\ref{sec:boundary} we introduce so-called boundary quantum
phase transitions -- those are transitions where only degrees of freedom of
a subsystem become critical.
This concept is illustrated using examples of quantum impurity problems,
which are of current interest both in correlated bulk materials and in
mesoscopic physics.
We end in Sec.~\ref{sec:sum} with a brief summary and outlook.
To shorten notations, we will often employ units such that $\hbar = k_B = 1$.


\section{Quantum phase transitions}
\label{sec:qpt}

This section will give a general introduction into both classical and quantum
phase transitions, and point out similarities and important conceptual
differences between the two.


\subsection{Basic concepts of phase transitions}

We start out with briefly collecting the basic concepts of phase transitions
and critical behavior \cite{Ma76,Goldenfeld92}
which are necessary for the later discussions.
Phase transitions are traditionally classified into first-order and continuous
transitions.
At first-order transitions the two phases co-exist at the transition temperature --
examples are ice and water at 0$^\circ$ C, or water and steam at 100$^\circ$ C.
In contrast, at continuous transitions the two phases do not co-exist.
An important example is the ferromagnetic transition of iron at 770$^\circ$ C,
above which the magnetic moment vanishes.
This phase transition occurs at a point where thermal fluctuations destroy
the regular ordering of magnetic moments -- this happens continuously in
the sense that the magnetization vanishes continuously when approaching the
transition from below.
The transition point of a continuous phase transition is also called critical
point.
The study of phase transitions, continuous phase transitions in particular,
has been one of the most fertile branches of theoretical physics in the
last decades.

In the following we concentrate on systems near a continuous phase transition.
Such a transition can usually be characterized by an order parameter --
this is a thermodynamic quantity that is zero in one phase
(the disordered) and non-zero and non-unique in the other (the ordered)
phase.
Very often the choice of an order parameter for a particular
transition is obvious as, e.g., for the ferromagnetic transition where
the total magnetization is an order parameter.
However, in some cases finding an appropriate order parameter is complicated
and still a matter of debate, e.g., for the interaction-driven metal--insulator
transition in electronic systems (the Mott transition \cite{Mott}).

While the thermodynamic average of the order parameter is zero in the
disordered phase, its fluctuations are non-zero. If the critical point
is approached, the spatial correlations of the
order parameter fluctuations become long-ranged. Close to the
critical point their typical length scale, the correlation length $\xi$,
diverges as
\begin{equation}
\xi \propto |t|^{-\nu}
\label{divxi}
\end{equation}
where $\nu$ is the correlation length critical exponent and
$t$ is some dimensionless measure of the distance from the critical point.
If the transition occurs at a non-zero temperature $T_c$,
it can be defined as $t=|T-T_c|/T_c$.
In addition to the long-range correlations in space there are
analogous long-range correlations of the order parameter
fluctuations in time. The typical time scale for a decay of
the fluctuations is the correlation (or equilibration) time
$\tau_c$. As the critical point is approached the correlation
time diverges as
\begin{equation}
  \tau_c \propto \xi^z \propto |t|^{-\nu z}
\label{divtau}
\end{equation}
where $z$ is the dynamic critical exponent.
Close to the critical point there is no characteristic
length scale other than $\xi$ and no characteristic time
scale other than $\tau_c$.
(Note that a microscopic
cutoff scale must be present to explain non-trivial critical
behavior, for details see, e.g., Goldenfeld \cite{Goldenfeld92}.
In a solid such a scale is, e.g., the lattice
spacing.)

The divergencies (\ref{divxi}) and (\ref{divtau}) are responsible
for the so-called critical phenomena.
At the phase transition point, correlation length and time are
infinite, fluctuations occur on {\em all} length and time scales,
and the system is said to be scale-invariant.
As a consequence, all observables depend via power laws on the external
parameters.
The set of corresponding exponents -- called critical exponents --
completely characterizes the critical behavior near a particular phase
transition.

Let us illustrate the important concept of scaling in more detail.
To be specific, consider a classical ferromagnet with the order parameter
being the magnetization $M({\bf r})$. External parameters are
the reduced temperature $t=|T-T_c|/T_c$ and the external
magnetic field $B$ conjugate to the order parameter.
Close to the critical point the correlation length is the only relevant
length scale, therefore the physical properties must be unchanged if
we rescale all lengths in the system by a common factor,
and at the same time adjust the external parameters in such a way
that the correlation length retains its old value.
This gives rise to the homogeneity relation for the
singular part of the free energy density,
\begin{equation}
 f(t,B) = b^{-d} f(t\, b^{1/\nu}, B\, b^{y_B}).
\label{eq:widom}
\end{equation}
Here $y_B$ is another critical exponent.
The scale factor $b$ is an arbitrary positive number.
Analogous homogeneity relations for other thermodynamic quantities
can be obtained by differentiating $f$. The homogeneity law
(\ref{eq:widom}) was first obtained phenomenologically by
Widom \cite{Widom65}; within the framework of the renormalization group
theory \cite{Wilson71} it can be derived from first principles.

In addition to the critical exponents $\nu,~ y_B$ and $z$ defined above,
a number of other exponents is in common use. They describe the
dependence of the order parameter and its correlations on the
distance from the
critical point and on the field conjugate to the order parameter.
The definitions of the most commonly used
critical exponents are summarized in Table \ref{table:exponents}.
\begin{table}
\caption{Commonly used critical exponents for magnets,
  where the order parameter is the magnetization $m$, and the conjugate field
  is a magnetic field $B$. $t$ denotes the
  distance from the critical point and $d$ is the space dimensionality.
  (The exponent $y_B$ defined in (\ref{eq:widom}) is related to $\delta$
  by $y_B=d \, \delta /(1+\delta)$.)}
\renewcommand{\arraystretch}{1.2}
\vspace{2mm}
\begin{tabular*}{\textwidth}{c@{\extracolsep\fill}ccc}
\hline
\hline
&exponent& definition & conditions \\
\hline
specific heat &$\alpha$& $C \propto |t|^{-\alpha}$ & $t \to 0, B=0$\\
order parameter& $\beta$ & $m \propto (-t)^\beta$ & $t \to 0$ from below, $B=0$\\
susceptibility& $\gamma$ & $\chi \propto |t|^{-\gamma}$ & $t \to 0, B=0$\\
critical isotherm & $\delta$ & $B \propto |m|^\delta {\rm sign}(m)$ & $B \to 0, t=0$\\
\hline
correlation length& $\nu$ & $\xi \propto |t|^{-\nu}$ & $t \to 0, B=0$\\
correlation function& $\eta$ & $G(r) \propto |r|^{-d+2-\eta}$ & $t=0, B=0$\\
\hline
dynamic& $z$ & $\tau_c \propto \xi^{z}$ & $t \to 0, B=0$\\
\hline
\hline
\end{tabular*}
\vspace*{2mm}
\label{table:exponents}
\end{table}
Note that not all the exponents defined in Table \ref{table:exponents}
are independent from each other.
The four thermodynamic exponents $\alpha, \beta,\gamma,\delta$ can
all be obtained from the free energy (\ref{eq:widom}) which contains
only two independent exponents.
They are therefore connected by the so-called scaling relations
\begin{eqnarray}
2- \alpha =  2 \beta +\gamma\,, ~~
2 - \alpha  = \beta ( \delta + 1)~.
\end{eqnarray}
Analogously, the exponents of the correlation length and correlation
function are connected by two so-called hyperscaling relations
\begin{eqnarray}
2- \alpha =  d\,\nu \, ,~~
\gamma = (2-\eta) \nu~.
\label{hyper}
\end{eqnarray}
(Hyperscaling relations are violated in theories with
mean-field critical behavior due to the presence of a
dangerously irrelevant variable, see Sec.~\ref{sec:dim}.)
Since statics and dynamics decouple in classical statistics (Sec.~\ref{sec:qucl})
the dynamic exponent $z$ is completely independent from all the others.

One of the most remarkable features of continuous phase transitions is
universality, i.e., the fact that the critical exponents are the same for
entire classes of phase transitions which may occur in very different
physical systems.
These universality classes are determined only
by the symmetries of the order parameter and by the space
dimensionality of the system.
This implies that the critical exponents of a phase transition occurring in
nature can be determined exactly (at least in principle) by investigating
{\em any} simple model system belonging to the same universality class.
The mechanism behind universality is again the divergence of the correlation
length. Close to the critical point the system effectively averages over
large volumes rendering the microscopic details of the Hamiltonian
unimportant.


\subsection{Quantum mechanics and the vicinity of the critical point}

The question of to what extent quantum mechanics is important
for understanding a continuous phase transition has at least two
aspects.
On the one hand, quantum mechanics can be essential to understand
the existence of the ordered phase, (e.g., superconductivity) -- this
depends on the particular transition considered.
On the other hand, one may ask whether quantum mechanics influences
the asymptotic critical behavior.
For this discussion we have to compare two energy scales, namely
$\hbar\omega_c$, which is the typical energy of long-distance order parameter
fluctuations, and the thermal energy $k_B T$.
We have seen in the preceeding section that the typical time scale $\tau_c$
of the fluctuations diverges as a continuous transition is approached.
Correspondingly, the typical frequency scale $\omega_c$ goes to zero and with it the typical
energy scale
\begin{equation}
  \hbar \omega_c \propto |t|^{\nu z}~.
  \label{eq:energy scale}
\end{equation}
Quantum mechanics will be important as long as this typical energy scale
is larger than the thermal energy $k_B T$; on the other hand, for
$\hbar\omega_c \ll k_B T$ a purely classical description can be
applied to the order parameter fluctuations.
In other words, the character of the order parameter fluctuations
crosses over from quantum to classical when $\hbar\omega_c$
falls below $k_B T$.

Now, for any transition occurring at some finite temperature $T_c$
quantum mechanics will become unimportant for $|t| < T_c^{1/\nu z}$,
in other words, the critical behavior asymptotically close to the
transition is entirely classical.
This justifies to call all finite-temperature phase transitions
``classical''.
Quantum mechanics can still be important on
microscopic scales, but classical thermal fluctuations
dominate on the macroscopic scales that control the critical behavior.
If, however, the transition occurs at zero temperature as a function of
a non-thermal parameter $r$ like pressure or magnetic field, the behavior
is always dominated by quantum fluctuations. Consequently,
transitions at zero temperature are called ``quantum'' phase transitions.

The interplay of classical and quantum fluctuations leads to
an interesting phase diagram in the vicinity of the quantum critical
point.
Two cases need to be distinguished, depending on whether long-range
order can exist at finite temperatures.

\begin{figure}[t]
\epsfxsize=15cm
\centerline{\epsffile{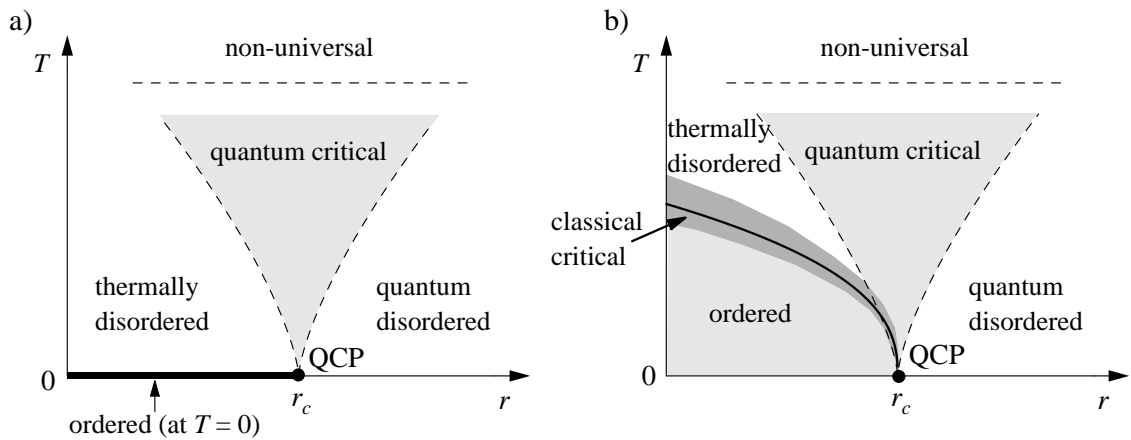}}
\vspace*{-0.3cm}
\caption[fig1]
{
Schematic phase diagrams in the vicinity of a quantum critical point (QCP).
The horizontal axis represents the control parameter $r$ used to tune
the system through the quantum phase transition, the vertical axis is
the temperature $T$.
a) Order is only present at zero temperature.
The dashed lines indicate the boundaries of the quantum critical region
where the leading critical singularities can be observed;
these crossover lines are given by $k_B T \propto |r-r_c|^{\nu z}$.
b)~Order can also exist at finite temperature.
The solid line marks the finite-temperature boundary between the ordered
and disordered phases. Close to this line, the critical behavior
is classical.
}
\label{fig:qcpvic}
\end{figure}

Fig.~\ref{fig:qcpvic}a describes the situation where order only exists at $T=0$, this is the case,
e.g., in two-dimensional magnets with SU(2) symmetry where order at finite $T$ is forbidden
by the Mermin-Wagner theorem.
In this case there will be no true phase
transition in any real experiment carried out at finite temperature.
However, the finite-$T$ behavior is
characterized by three very different regimes, separated by crossovers, depending on whether the
behavior is dominated by thermal or quantum fluctuations of the order parameter.
In the thermally disordered region the long-range order is destroyed mainly by
thermal order parameter fluctuations.
In contrast, in the quantum disordered region the physics is
dominated by quantum fluctuations, the system
essentially looks as in its quantum disordered ground state at $r>r_c$.
In between is the so-called
quantum critical region \cite{CHN}, where both types of fluctuations are important.
It is located near the critical parameter value $r=r_c$ at comparatively
high (!) temperatures.
Its boundaries are determined by the
condition $k_B T > \hbar \omega_c \propto |r-r_c|^{\nu z}$: the system
``looks critical'' with respect to the tuning parameter $r$, but is driven
away from criticality by thermal fluctuations.
Thus, the physics in the quantum critical region is controlled by the thermal excitations of the
quantum critical ground state, whose main characteristics is the {\em absence}
of conventional quasiparticle-like excitations.
This causes unusual finite-temperature properties in the quantum critical
region, such as unconventional power laws, non-Fermi liquid behavior etc.
Universal behavior is only observable in the vicinity of the quantum critical
point, i.e., when the correlation length is much larger than microscopic length
scales.
Quantum critical behavior is thus cut off
at high temperatures when $k_B T$ exceeds characteristic microscopic energy
scales of the problem -- in magnets this cutoff is, e.g., set by the typical
exchange energy.

If order also exists at finite temperatures, Fig.~\ref{fig:qcpvic}b,
the phase diagram is even richer.
Here, a real phase transition is encountered upon variation of $r$
at low $T$;
the quantum critical point can be viewed as the endpoint of a line of
finite-temperature transitions.
As discussed above, classical fluctuations will dominate in the vicinity of
the finite-$T$ phase boundary, but
this region becomes narrower with decreasing temperature, such that
it might even be unobservable in a low-$T$ experiment.
The fascinating quantum critical region is again at finite temperatures above
the quantum critical point.

A quantum critical point can be generically approached in two different
ways: $r\to r_c$ at $T=0$ or $T\to 0$ at $r=r_c$.
The power-law behavior of physical observables in both cases can often be
related. Let us discuss this idea by looking at the entropy $S$.
It goes to zero at the quantum critical point (exceptions are impurity
transitions discussed in Sec.~\ref{sec:boundary}), but its derivatives
are singular.
The specific heat $C=T \partial S/\partial T$ is expected to show power-law
behavior, as does the quantity $B= \partial S/\partial r$.
Using scaling arguments (see below) one can now analyze the ratio of the
{\em singular parts} of $B$ and $C$: the scaling dimensions of $T$ and $S$ cancel,
and therefore $B/C$ scales as the inverse of the tuning parameter $r$.
Thus, one obtains a {\em universal} divergence in the low-temperature limit,
$B/C \propto |r-r_c|^{-1}$; similarly, $B/C \propto T^{-1/\nu z}$ at $r=r_c$ \cite{markus}.
Note that $B/C$ does not diverge at a finite-temperature phase transition.
At a pressure-tuned phase transition $r\equiv p$, then $B$ measures the thermal
expansion, and $B/C$ is the so-called Gr\"uneisen parameter.
As will be discussed in Sec.~\ref{sec:dim}, the scaling argument presented here can
be invalid above the upper-critical dimension.


\subsection{Quantum-classical mapping and scaling}
\label{sec:qucl}

To gain a deeper understanding of the relation between classical
and quantum behavior, and the possible quantum--classical crossover,
we have to recall general features from quantum statistical
mechanics.

The starting point for the derivation of thermodynamic
properties is the partition function
\begin{equation}
Z = {\rm Tr}\,e^{-H/k_B T}
\end{equation}
where $H = H_{\rm kin} + H_{\rm pot}$ is the Hamiltonian characterizing
the system.
In a classical system, the kinetic and potential part of $H$ commute,
thus $Z$ factorizes, $Z = Z_{\rm kin} Z_{\rm pot}$, indicating that in a
classical system statics and dynamics decouple.
The kinetic contribution to the free energy will usually not display any
singularities, since it derives from the product of simple Gaussian integrals.
Therefore one can study classical phase transitions using effective
time-independent theories, which naturally live in $d$ dimensions.

In contrast, in a quantum problem the kinetic and potential parts of $H$
in general do not commute, the quantum mechanical partition function
does {\em not} factorize, which implies that statics and dynamics
are always coupled.
An order parameter field theory needs to be formulated in terms of space
and time dependent fields.
The canonical density operator $e^{-H/k_B T}$ looks
exactly like a time evolution operator in imaginary time
$\tau$ if one identifies $1/k_B T = \tau = -i\Theta /\hbar$
where $\Theta$ denotes the real time.
Therefore it proves convenient to introduce an imaginary time direction
into the system -- formally this is done in the path integral representation
of the partition function.
At zero temperature the imaginary time acts similarly to an additional
space dimension since the extension of the system in this direction
is infinite. According to (\ref{divtau}), time scales like
the $z$-th power of a length.
(Note that $z\!=\!1$ for many transitions in clean insulators, however,
in general other values of $z$ including fractional ones can occur.)
The classical homogeneity law (\ref{eq:widom}) for the free energy density can
now easily be adopted to the case of a quantum phase transition.
At zero temperature it reads
\begin{equation}
 f(t,B) = b^{-(d+z)} f(t\, b^{1/\nu},B\, b^{y_B})
\label{eq:quantum widom}
\end{equation}
where now $t=|r-r_c|/r_c$.
This shows that a quantum phase transition in $d$ space dimensions is
related to a classical transition in $(d\!+\!z)$ space dimensions.

For a quantum system, it is now interesting to discuss the quantum to
classical crossover upon approaching a {\em finite-temperature} phase
transition -- this turns out to be equivalent to a dimensional
crossover.
Under the quantum--classical mapping the temperature of the quantum problem
maps onto the inverse length of the imaginary time axis.
(Note that the temperature of the classical problem corresponds to
a coupling constant in the quantum model within this mapping.)
Close to the transition (Fig.~\ref{fig:qcpvic}b) the behavior is determined by
the relation between the characteristic correlation time, $\tau_c$, and
the extension in imaginary time direction, $\beta=1/k_B T$.
The crossover from quantum to classical behavior will occur when the
correlation time $\tau_c$ becomes larger than $\beta$ which is equivalent
to the condition $|t|^{\nu z} < k_B T$;
in other words, once $\tau_c > \beta$ the system realizes that it is
effectively only $d$-dimensional and not $(d\!+\!z)$-dimensional.
The corresponding crossover scaling is equivalent to finite size scaling in
imaginary time direction.

In contrast, when approaching a {\em zero-temperature} transition by lowering the
temperature at $r=r_c$, both $1/k_B T$ and $\tau_c$ diverge, such that quantum
effects are always important, and we get a truly $(d\!+\!z)$-dimensional
system.
Recognizing that quantum critical singularities can be cut off both
by tuning $r$ away from $r_c$ at $T=0$ and by raising $T$ at $r=r_c$,
it is useful to generalize the homogeneity law (\ref{eq:quantum widom}) to
finite temperatures,
\begin{equation}
 f(t,B,T) = b^{-(d+z)} f(t\, b^{1/\nu},B\, b^{y_B}, T \, b^z)~.
\label{eq:temperature quantum widom}
\end{equation}
The homogeneity law for the free energy immediately leads to
scaling behavior of both static and dynamic observables.
Consider an observable ${\cal O}(k,\omega)$, e.g., a magnetic
susceptibility, measured at wavevector $k$ and frequency $\omega$,
then the existence
of a single length scale $\xi$ and a single time scale $\omega_c^{-1} = \xi^z$
implies
\begin{eqnarray}
{\cal O}(t,k,\omega,T)
&=& \xi^{d_{\cal O}}  ~  O_1(k \xi, \omega \xi^z, T \xi^z)  \nonumber\\
&=& T^{-d_{\cal O}/z} ~  O_2(k T^{-1/z}, \omega / T, T \xi^z)
\label{obs_scal}
\end{eqnarray}
where $O_1$ and $O_2$ are different forms for the scaling function associated
with the observable $\cal O$, and ${d_{\cal O}}$ is the so-called scaling
dimension of $\cal O$.
For simplicity we have set the external field $B=0$ and
assumed that $k$ is measured relative to the ordering wavevector.

Precisely at the critical point, where the correlation length is infinite,
the only length is set by the measurement wavevector $k$, analogously, the
only energy is $\omega$, leading to
\begin{equation}
{\cal O}(t=0,k,\omega,T=0)
= k^{-d_{\cal O}} ~ O_3(k^z / \omega) \,.
\label{obs_scalqcp}
\end{equation}
Similarly, for $k=0$, but a finite temperatures, we obtain
\begin{equation}
{\cal O}(t=0,k=0,\omega,T)
= T^{-d_{\cal O}/z} ~ O_4(\omega/T) \,,
\label{woverT}
\end{equation}
this is known as ``$\omega/T$'' scaling.
It is important to note that the relations (\ref{obs_scal})--(\ref{woverT}),
sometimes also called ``naive scaling'',
are only expected to be valid if the critical point satisfies hyperscaling properties,
which is true below the upper-critical dimension, see Sec.~\ref{sec:dim}.

The power of the quantum--classical analogy can be nicely demonstrated
by considering a correlation function at the critical point.
In a classical $d$-dimensional system, correlations typically
fall off with a power law in real space implying a momentum
space behavior as
\begin{equation}
G(k) \propto k^{-2 + \eta_d} ~.
\label{corr_1}
\end{equation}
If we are now interested in a quantum phase transition with
$z=1$ which maps onto a $(d\!+\!1)$-dimensional classical theory
with known properties, we can conclude that at $T=0$
\begin{equation}
G(k,i\omega_n) \propto [k^2+\omega_n^2]^{(-2 + \eta_{d+1})/2}
\label{corr_2}
\end{equation}
where the Matsubara frequency $\omega_n$ simply takes the
role of an additional wavevector component.
For real frequencies, the above translates into a retarded Green's function
of the form
\begin{equation}
G^R(k,\omega) \propto [k^2-(\omega+i\delta)^2]^{(-2 + \eta_{d+1})/2} ~.
\label{corr_3}
\end{equation}
This examples illustrates the character of the excitation spectrum
at a quantum critical point: $G^R$ does not show a conventional
quasiparticle pole, but instead a branch cut for $\omega > k$,
corresponding to a critical continuum of excitations.
This implies that modes become overdamped,
and the system shows quantum relaxational dynamics \cite{book}.

Now, if a quantum transition seems to map generically onto a classical
transition, what is different about quantum transitions?
For a number of reasons it turns out that {\em not}
all needed properties of a given quantum system can be obtained
from a classical theory, the reasons being:

(A) Theories for quantum systems can have ingredients which
make them qualitatively different from their classical counterparts;
examples are topological Berry phase terms and
long-ranged effective interactions arising from soft modes.

(B) Even if a quantum--classical mapping is possible, calculating real time dynamics
of the quantum system often requires careful considerations and novel theories,
as an approximation done on the imaginary time axis is in general not
appropriate for real times~\cite{book}.

(C) Quenched disorder leads to an extreme anisotropy in
space-time in the classical theory~\cite{bk01},
see Sec.~\ref{sec:disorder}.

(D) The dynamic properties of the quantum system near a quantum
critical point are characterized by a fundamental new time scale, the
{\em phase coherence time} $\tau_\phi$, which has no analogue at the corresponding classical
transition~\cite{book}.
Notably, $\tau_\phi$ diverges as $T\to 0$ for all parameter values,
i.e., the quantum system has perfect phase coherence even in the disordered
phase -- this seems to be peculiar considering that all correlations
decay exponentially in the high-temperature phase of the corresponding
$(d\!+\!z)$-dimensional classical model.
Technically, this peculiarity is related to the nature of
the analytic continuation between imaginary and real times.
In many models for quantum phase transitions, it is found that
$\tau_\phi \propto 1/T$ as $T\to 0$ at the critical coupling,
but $\tau_\phi$ diverges faster in the stable phases.

Parenthetically, we note that ``conventional'' quantum criticality leads to a power law behavior
in dynamic observables as function of energy and momentum, where (at $T\!=\!0$) the
singularity is cut off both by finite frequency and finite momentum
(measured from the ordering wavevector).
There are, however, interesting cases where the order parameter is not
associated with a particular momentum, and then so-called ``local'' critical
behavior may occur.
Candidate transitions are those involving spin glass or topological
order; interesting recent proposals involve the breakdown of Kondo
screening in heavy fermion compounds (Sec.~\ref{sec:metals}).


\subsection{Ground state properties near a quantum phase transition}

It is useful to think about a quantum transition from
the point of view of many-body eigenstates of the system.
A quantum phase transition is a non-analyticity of the ground state
properties of the system as function of the control parameter $r$.
If this singularity arises from a simple {\em level crossing} in the
many-body ground state, then we have a first-order quantum phase transition,
without diverging correlations and associated critical
singularities.
A first-order quantum transition can also occur in a finite-size system.
The situation is different for continuous transitions, where a higher-order
singularity in the ground state energy occurs:
here an infinite number of many-body eigenstates are involved,
and the thermodynamic limit is required to obtain a sharp transition.
For any finite-size system a continuous transition will be rounded into
a crossover --
this is nothing but an avoided level crossing in the ground state, which
then becomes infinitely sharp in the thermodynamic limit.

For numerical simulations of finite-size systems the choice of boundary conditions
plays an important role: if the infinite system shows
a first-order transition where the two ground states can be distinguished
by a certain quantum number, then such a transition will also be rounded into
a crossover for finite system size if the boundary conditions spoil
the existence of this quantum number.
This applies, e.g., to simulations using open boundary conditions
which invalidate momentum or a certain parity as good quantum numbers.


\subsection{Order parameters and order parameter field theories}
\label{sec:fieldth}

Near a phase transition the relevant physics is described by
long-wavelength order parameter fluctuations.
Therefore it appears appropriate to disregard microscopic processes and to
work with a theory containing the order parameter fluctuations only.
However, in itinerant electron systems
it can be neccessary to include low-energy fermionic excitations explicitly
in the critical theory~\cite{bkv} because of the strong coupling
between order parameter and fermionic particle-hole modes,
see Secs.~\ref{sec:metals} and \ref{sec:sc}.

Formally, an effective theory can be obtained in the path integral
formulation from a microscopic model of interacting particles
(e.g. fermions).
To this end, the interaction terms are suitably decoupled
via a Hubbard-Stratonovich transformation, and then the microscopic degrees
of freedom can be integrated out, resulting in a theory of the
order parameter fluctuations alone.
Near a phase transition, the correlation length is large, and therefore the
effective theory can be formulated in the continuum limit,
with a proper ultraviolet cutoff set by the lattice spacing.
In most cases the resulting theory will contain simple powers of the
order parameter field and its gradients,
and the form of such a theory can actually be derived based on symmetry
arguments only.
The form of the low-energy effective action will thus only depend on
the dimensionality and symmetry of the underlying system and the
symmetry of the order parameter -- reflecting nothing but the celebrated
universality of critical behavior.

In the following we mention a few examples of models and corresponding
quantum field theories; concrete applications will be given
in Sec.~\ref{sec:bulk}.
We will not discuss here the special theories appropriate for
one-dimensional systems, the reader is referred to Ref.~\cite{book}
for an introduction.

\subsubsection{Ising and rotor models.}

Quantum rotor models, with the Hamiltonian
\begin{equation}
H_R = \frac{J g}{2} \sum_i \hat{\bf L}_i^2 -
J \sum_{\langle ij\rangle} \hat{\bf n}_i \cdot \hat{\bf n}_j \,,
\label{rotor}
\end{equation}
are some of the simplest models showing quantum phase transitions.
Here, the $N$-component operators $\hat{\bf n}_i$ with $N\geq 2$
represent the orientation of rotors
on sites $i$ of a regular $d$-dimensional lattice, with $\hat{\bf n}_i^2=1$,
and the $\hat{\bf L}_i$ are the corresponding angular momenta.
The nearest-neighbor coupling in the second term prefers a ferromagnetic ordering,
whereas the kinetic term tends to delocalize the individual rotors.
Thus, $g \ll 1$ leads to an ordered state, whereas $g \gg 1$ places the
system into a quantum-disordered phase~\cite{book}.
A related model with $N\!=\!1$ components is the quantum Ising model
discussed in more detail in Sec.~\ref{sec:lihof}.

The order parameter describing the transition in the model (\ref{rotor})
is the ferromagnetic magnetization;
a continuum theory can be derived from the lattice model in the
path integral formulation by spatial {\em coarse-graining}, i.e.,
the variables $\hat{\bf n}_i$ are averaged over microscopic length
scales.
The magnitude of the continuum order parameter field, $\bf\phi$, can thus vary over a wide range,
although the ``magnitude'' of the underlying microscopic object
(a single rotor) is fixed.
Expanding the resulting action into lowest-order field gradients,
one arrives at the so-called $\phi^4$ theory.
This quantum field theory, described by the action
\begin{eqnarray}
Z &=& \int {\cal D} \phi_\alpha (x,\tau) ~ \exp(-{\cal S}) \,, \nonumber \\
{\cal S} &=&
\int d^d x \int_0^{\hbar/k_B T}\!\!\!\!\! d \tau \bigg[
\frac{1}{2} (\partial_{\tau} \phi_{\alpha})^2 + \frac{c^2}{2} ( \nabla_{x}
\phi_{\alpha} )^2  + \frac{r}{2} \, \phi_{\alpha}^2
+ \frac{g_0}{4!} \left( \phi_{\alpha}^2 \right)^2 \bigg]
\label{phi4} ~,
\end{eqnarray}
turns out to be a generic theory, which we will refer to frequently in
this article, being appropriate for systems where the $N$-component order parameter
field, $\phi_\alpha(x,\tau)$, is a non-conserved density.
$r$ is the bare ``mass'' of the order parameter fluctuations,
used to tune the system across the quantum phase transition,
and $c$ is a velocity.
$g_0$ describes the important self-interactions of the $\phi$
fluctuations, and we will comment on them in the next section.
Note that the dynamic exponent is $z=1$, i.e., space and time enter
symmetrically.
Although the order parameter fluctuations are bosonic in nature,
the second-order time derivative in (\ref{phi4}) does not represent
the dynamics of a canonical boson, but rather arises from a gradient
expansion of the order parameter action.

The model Eq. (\ref{phi4}) obeys a O($N$) symmetry, corresponding to
rotations in order parameter space.
$N=1$ corresponds to a scalar or Ising order parameter;
for $N\geq 2$ the order parameter represents a vector
for which one can distinguish longitudinal (amplitude) and
transverse (phase or direction) fluctuations.
For $N=3$, the theory can describe spin degrees of freedom
in an insulator; in a number of cases, however, Berry phases
are important which we will explicitely discuss below.
In writing down (\ref{phi4}) we have assumed $\phi$ to be real.
Situations with complex $\phi$ can appear as well;
then the action can contain two non-linear terms,
$(|\phi_\alpha|^2)^2$ and $|\phi_\alpha^2|^2$.
We will see in Sec.~\ref{sec:OP} that the phase of a complex order parameter field
usually corresponds to a sliding degree of freedom of a density
wave.

At positive values of $r$, the model (\ref{phi4}) is in a disordered phase,
characterized by $\langle \phi \rangle = 0$;
at $T=0$ the propagator of the $\phi$ fluctuations has a
gap of size $\Delta(T=0) = r$ (to zeroth order in $g_0$).
With decreasing $r$, this gap decreases, vanishing at the
quantum phase transition to the ordered phase.
In the ordered phase, the magnitude of the order parameter
is determined by $g_0$, at $T=0$ at the mean-field level it is
$\langle \phi \rangle = -r / (2 g_0)$.
For $N\geq 2$ the ordered phase breaks a continuous symmetry and
supports gapless Goldstone modes with linear dispersion.

Near the critical point, for $N\geq 2$ both amplitude and phase
fluctuations can be important.
However, as will be discussed in Secs.~\ref{sec:dim} and \ref{sec:RG},
below the so-called upper-critical dimension the critical behavior is
determined by phase fluctuations, and in this case a theory
which neglects amplitude fluctuation -- a ``hard-spin'' theory
in contrast to the ``soft-spin'' formulation of the
$\phi^4$ model -- is adequate.
The hard-spin theory is formulated in terms of a field $n(x,\tau)$
with a unit-length constraint and is known as the
non-linear sigma model:
\begin{eqnarray}
Z &=& \int {\cal D} {\bf n}(x,\tau) ~ \delta({\bf n}^2-1) ~ \exp(-{\cal S}) \,,\nonumber \\
{\cal S} &=& \frac{N}{2 c g_0}
\int d^d x \int_0^{\hbar/k_B T}\!\!\!\!\! d \tau \bigg[
(\partial_{\tau} n_{\alpha})^2 + c^2 ( \nabla_{x}
n_{\alpha} )^2 \bigg]
\label{nlsm} ~.
\end{eqnarray}
For dimensions $1<d<3$, equivalently $2<D<4$ with $D=d+z$,
the $\phi^4$ model (\ref{phi4}) and the non-linear sigma
model (\ref{nlsm}) represent different approaches to describe the
same quantum critical point, and their critical properties are
identical \cite{bz}.
Technically, the $\phi^4$ model can be analyzed in an expansion
about the upper-critical dimension, i.e., for $D=4-\epsilon$,
whereas the non-linear sigma model allows for an expansion about
the lower-critical dimension $D=2+\epsilon$ -- those will be
briefly discussed in Sec.~\ref{sec:RG}.

Both models (\ref{phi4}) and (\ref{nlsm}) possess propagating
modes which are undamped at low energies, and the quantum transition is characterized
by $z=1$.
In systems with gapless fermionic excitations, order parameter
fluctuations can, however, decay into particle-hole pairs, and this damping
requires modifications of the described theories (or even different
approaches), see Sec.~\ref{sec:fermions}.

As usual, theories of the type (\ref{phi4}), supplemented by
higher-order terms and/or suitable coupling to external
fields, can also describe first-order transitions.
There, the transition point is characterized by the existence of two
distinct degenerate global minima in the free-energy landscape,
which usually leads to a jump in the order parameter and other observables
when tuning the system through the transition.
However, no universality is present at first-order transitions, and
a continuum description is only appropriate for transitions being weakly
first order.

\subsubsection{Heisenberg spins.}

Deriving a continuum model for a system of localized quantum-mechanical
spins can be conveniently done using spin-coherent state path integrals,
and the reader is referred to Ref.~\cite{book} for an introduction.
The orientation of each spin is described by a
three-component unit vector $\bf N$, and coherent states $|{\bf N}\rangle$ obey
$\langle{\bf N}|{\hat S}|{\bf N}\rangle = S{\bf N}$.
The dynamics of a single spin takes the form of a Berry phase
term, ${\cal S}_B = \int d\tau \langle{\bf N}|\partial_\tau|{\bf N}\rangle$,
which is purely imaginary.
This term can be shown to be equal to $i S$ times
the solid angle subtended by the spin vector on the unit sphere
(note ${\bf N}(\beta) = {\bf N}(0)$) -- importantly this Berry
phase term contains a first-order time derivative.

The order parameter of interest is again a magnetization density, e.g.,
for antiferromagnets a staggered magnetization.
Starting from a lattice spin model, a continuum action can
be derived which contains gradients of the order parameter field, e.g.,
of the form (\ref{phi4}) or (\ref{nlsm}),
plus a term arising from the spin Berry phases \cite{book}.
The behavior of resulting low-energy theory crucially depends on those
Berry phases, and is completely different for ferromagnets and
antiferromagnets:
In ferromagnetic spin systems the Berry phase terms add up and determine
the dynamics of the system, as they dominate over the second-order
time derivative in (\ref{phi4}) or (\ref{nlsm}) --
this is related to the fact that the order parameter is a {\em conserved} density
here.
In contrast, in antiferromagnets the Berry phase contributions
oscillate in sign from site to site, and on average simply
cancel.
Therefore, in many cases Berry phases can essentially be neglected,
e.g., in clean antiferromagnets on high-dimensional regular lattices,
and the resulting model is a $\phi^4$ or non-linear sigma model.
However, {\em singular} Berry phase configurations can play a role.
This is crucial in one space dimension, where a corresponding
topological term appears in the action for half-integer spins,
which leads to the well-known gapless ``critical'' phase with
power law decay of spin correlations for the $d\!=\!1$ Heisenberg model.
In contrast, for integer spins the topological term is absent,
and the system is characterized by an excitation gap and
conventional short-range spin correlations.

In a number of cases the analysis of the spin model can be carried
out in terms of auxiliary particles.
One example is the Dyson-Maleev representation of spin operators in
terms of bosons;
another one is the bond-boson representation for spin pairs (dimers) in
quantum paramagnets \cite{book}.
In both situations one arrives at theories similar to Bose gases which will
be discussed below.

\subsubsection{Boson and fermion models.}

Microscopic models of bosons and fermions can have a plethora of ordered
phases.
In many cases, the order parameter can be represented as a boson or
fermion bilinear, like for spin or charge density waves, and
the critical theory is of $\phi^4$-type with proper modifications
(Sec.~\ref{sec:metals}).

In addition, models of bosons or fermions with a conserved charge $Q$
can have a transition between a phase where $\langle Q\rangle$ is
pinned to a quantized value, and a phase where $Q$ varies smoothly
as function of external parameters.
In those cases, the density is an appropriate order parameter;
however, the theory has to be formulated in terms of Bose or Fermi fields
and will contain imaginary Berry phase terms for the particle dynamics.
For instance, varying the chemical potential through zero drives a quantum
phase transition where the zero-temperature particle density is zero in one
phase and finite in the other.
For bosons in $d>2$ this dilute gas quantum critical point is the endpoint of a line
of finite-$T$ transitions, associated with Bose-Einstein
condensation \cite{book}.

Interesting physics obtains in the presence of strong local interactions,
the Hubbard model being a paradigmatic example.
The physics of the fermionic Hubbard model is a subject of vast active research,
and cannot be reviewed here (see Sec.~\ref{sec:mit} for some
remarks on the Mott-Hubbard metal--insulator transition).
The bosonic Hubbard model is somewhat simpler; it was originally introduced
to describe spinless bosons on a lattice, representing Cooper pairs
undergoing Josephson tunneling between superconducting islands,
with a strong on-site Coulomb repulsion.
This model shows superfluid and insulating phases, and an appropriate
order parameter is the expectation value of a bosonic field,
representing the superfluid density (see Sec.~\ref{sec:sit}).

Microscopic realizations of hardcore boson models also arise from models of
quantum spins, where the degrees of freedom of spins or spin clusters can be
re-written in terms of bosons with infinite on-site repulsion.
An nice example of a dilute Bose gas critical point described above is found
in quantum antiferromagnets in an external field, where the zero-field ground
state is a gapped singlet state:
Increasing field splits the triplet excitation energies, and at a critical
field $H_c$ the lowest triplet condenses, leading to a state with transverse
long-range order.
Both the $T=0$ transition and the finite-$T$ transition (present in $d>2$),
where the transverse order is established, can be understood as ``Bose-Einstein
condensation of magnons''.
Interestingly, amplitude and phase of the condensate wavefunction correspond
to magnitude and direction of the transverse magnetization, and can thus
be measured directly.


\subsection{Fluctuations and critical dimensions}
\label{sec:dim}

The critical behavior at a particular transition is crucially determined
by the role played by {\em fluctuations} around the mean-field
solution, and their mutual interactions, e.g., the $g_0 \phi^4$ term
in the $\phi^4$ theory.
It turns out that fluctuations become increasingly important if the
space dimensionality of the system is reduced. Above a certain
dimension, called the upper-critical dimension $d_c^+$, fluctuations are
irrelevant, and the critical behavior is identical to that predicted by
mean-field theory (Ginzburg criterion).
For classical systems with short-range interactions and a scalar or vector order
parameter one finds $d_c^+=4$;
the mean-field exponents of the corresponding $\phi^4$ theory are:
\begin{equation}
\alpha=0\,,~
\beta=1/2 \,,~
\gamma=1\,,~
\delta=3\,,~
\nu=1/2\,,~
\eta = 0\,.
\end{equation}
Between $d_c^+$ and a second, smaller special dimension, called
the lower-critical dimension $d_c^-$, a phase transition exists, but
the critical behavior is different from mean-field theory.
Below $d_c^-$ fluctuations become so strong that they completely suppress
the ordered phase; we have $d_c^-=2$ for the classical $\phi^4$ model
with $N>1$ components.

The quantum--classical mapping discussed above implies that
for a quantum phase transition the critical dimensions are reduced
by $z$ compared to the corresponding classical transition.
Thus, in $d\!=\!3$ many quantum transitions show mean-field
behavior (supplemented by logarithmic corrections if $z=1$).

Technically, the existence of the upper-critical dimension is
connected with the role played by the ultraviolet cutoff of the
quantum field theory.
For $d<d_c^+$ the integrals arising in perturbation theory are
ultraviolet convergent, implying that the cutoff can be sent to
infinity.
Thus, the critical behavior becomes cutoff-independent, i.e., truly universal,
and observables will follow ``naive'' scaling, including ``$\omega/T$'' scaling
in dynamical quantities, as exploited in Sec.~\ref{sec:qucl},
and exponents will fulfill hyperscaling relations (\ref{hyper}).
Physically, this universality arises from the fact that the non-linear interaction
terms, e.g., $g$ in the $\phi^4$ theory, flow towards universal values
in the low-energy limit.
In contrast, for $d>d_c^+$, quantities will depend on the ultraviolet cutoff,
and non-universal corrections to observables arise from the non-linearities of
the original theory -- those are termed dangerously irrelevant.
Critical exponents will be independent of $d$ and thus hyperscaling is
violated.

Importantly, at the critical point in $d<d_c^+$ the integrals arising in
perturbation theory are {\em infrared} divergent. In other words,
(bare) perturbation theory itself is divergent below $d_c^+$ at $T\!=\!0$.
The trick to overcome this problem is the $\epsilon$ expansion described in the
next section.

For transitions which cannot be easily described by a local order parameter
field theory, the situation is less clear.
For instance, for the disorder-driven metal--insulator transition
(Sec.~\ref{sec:mit}) the lower-critical dimension is $d_c^-=2$;
it is likely that an upper-critical dimension does not exist \cite{bkmott}.


\subsection{Renormalization group approach and calculation of observables}
\label{sec:RG}

A powerful tool for the analysis of order-parameter field theories
is the renormalization group (RG) approach \cite{Wilson71}, which is
designed to determine the asymptotic low-energy, long-wavelength
behavior of a system.
The momentum-shell formulation of the RG proceeds by successively eliminating
high-energy degrees of freedom, and at the same time renormalizing the
couplings of the theory to keep physical properties invariant.
Doing so, the coupling constants flow as a function of the cutoff energy,
which is described by differential RG equations.
In the low-energy limit, the RG flow reaches a fixed point, and the corresponding
fixed point values of the couplings can be trivial (zero or infinity) or non-trivial.
The trivial values describe stable phases of the system,
whereas non-trivial finite values correspond to so-called intermediate-coupling
fixed points, which usually are critical fixed points describing continuous phase
transitions.

Analytical RG computations require a perturbative treatment of the interaction
terms of the field theory.
For $d<d_c^+$ these grow under RG transformations, i.e., interactions are
{\em relevant} in the RG sense below $d_c^+$.
For the $\phi^4$ theory (\ref{phi4}) the critical fixed point with
non-linearity $g=0$ (the so-called Gaussian fixed point) is unstable w.r.t. finite $g$,
i.e., the critical fixed point for $d<d_c^+$ will be characterized by a finite value
of $g$ (Wilson-Fisher fixed point).
The relevance of the interaction implies that bare perturbation theory is
divergent at criticality.
This problem can be overcome using the so-called $\epsilon$ expansion:
A theory of the $\phi^4$-type can be analyzed in an expansion around the
upper-critical dimension, i.e., in $\epsilon = d_c^+ -d$.
This approach is based on the observation that the fixed point value
of renormalized coupling $g$
is small near $d = d_c^+$, and a {\em double expansion} in
$g$ and $\epsilon$ allows for controlled calculations.
The RG differential equations are used to determine
RG fixed points and the corresponding values of the couplings.
For example, in the $\phi^4$ theory at criticality, the RG equation,
describing the flow of the dimensionless non-linear coupling $g$ upon reducing the
cutoff $\Lambda$, reads
\begin{equation}
\frac{d g}{d \ln \Lambda} \equiv \beta(g) = -\epsilon g + \frac{N+8}{6} g^2 + {\cal O}(g^3)
\end{equation}
with a fixed point at $g^\ast = (6\epsilon)/(N+8)$, where $\epsilon = d_c^+ -d$.
Physicswise, the finite value of $g$ at criticality implies strong self-interactions
of the order parameter bosons for $d<d_c^+$;
furthermore amplitude fluctuations of the order parameter in a $N\geq 2$ $\phi^4$ theory
are frozen out, and the critical dynamics is carried by phase
fluctuations.
A different expansion can be applied to the non-linear sigma model (\ref{nlsm}),
namely an expansion around the lower-critical dimension $d_c^-$, i.e.,
the expansion parameter is $\epsilon = d-d_c^-$.
Here, the underlying idea is that the transition temperature $t$ is small
near $d = d_c^-$, and the double expansion is done in $t$ and $\epsilon$
around the ordered state.
We note that $\epsilon$ expansions are strictly asymptotic expansions,
i.e., the limit $\epsilon\to 0$ has to be taken first, and convergence is
not guaranteed for any finite $\epsilon$.

Observables at and near criticality require different treatments
depending on whether $d > d_c^+$ or $d < d_c^+$.
Above the upper-critical dimension, bare perturbation is usually sufficient.
Below $d_c^+$, one can employ a renormalized perturbation expansion:
perturbation theory is formulated in terms of renormalized quantities,
in the final expressions the couplings are replaced by their fixed point values,
and the results are {\em interpreted} as arising from a $\epsilon$ expansion
of the expected power-law behavior, i.e.,
power laws are obtained by re-exponentiating the perturbation series.
The renormalized perturbation expansion can thus be understood as a certain
resummation technique of bare perturbation theory.
In some cases, a $1/N$ expansion, where $N$ is the number of order
parameter components, can also be used to calculate critical properties
for arbitrary $d$.

The calculation of dynamic and transport properties in the quantum critical regime
at finite temperatures turns out to be significantly more complicated.
Below $d_c^+$, dynamic properties are dominated by relaxation processes at frequencies
$\omega \ll T$, and in this regime the $\epsilon$ expansion fails.
However, considerable progress can be made by applying quasiclassical
techniques, where the excitations above the ground state can be
described by quasiclassical waves or particles \cite{book}.
For transport calculations, quasiclassical methods are not sufficient,
and a quantum-mechanical treatment using, e.g., a full quantum Boltzmann
equation is necessary. These techniques have been applied
to a number of models \cite{book}, but many issues including the
influence of disorder remain subject for future work.
In contrast, above $d_c^+$ perturbation theory is useful in many cases, and the
presence of quasiparticles allows to analyze transport using a
quasiclassical Boltzmann equation.

Beyond perturbative approaches, various numerical techniques have been
employed to investigate quantum phase transitions, some of which closely
follow the general idea of renormalization. One example is the
numerical renormalization group method (NRG) for quantum impurity
problems, proposed by Wilson \cite{nrg}.


\subsection{Example: The transverse-field Ising model}
\label{sec:lihof}

In this section we want to illustrate the general ideas presented above
by discussing the Ising model in a transverse field, which displays a paradigmatic
example of a quantum phase transition.
An experimental realization of this model can be found in the
low-temperature magnetic properties of LiHoF$_4$.
This material is an ionic crystal, and at sufficiently
low temperatures the only magnetic degrees of freedom are the spins
of the Holmium atoms. They have an easy axis, i.e., they prefer to point
up or down with respect to a certain crystal axis. Therefore they can
be represented by Ising spin variables. Spins at different
Holmium atoms interact via a dipolar interaction.
Without external magnetic field the ground state is a fully polarized
ferromagnet.

\begin{figure}[t]
\epsfxsize=8cm
\centerline{\epsffile{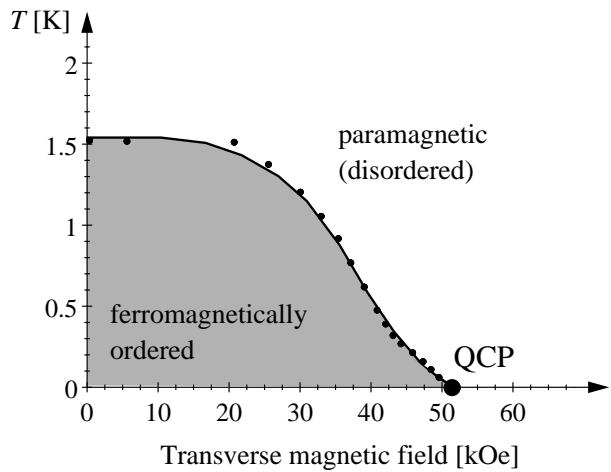}}
\caption{
Magnetic phase diagram of LiHoF$_4$ (after \protect\cite{BRA96}).
The ordered phase breaks a $Z_2$ symmetry of the Hamiltonian
and is bounded by a line of finite-temperature phase transitions.
This line terminates in the quantum critical point (QCP), where
ferromagnetic order is destroyed solely by quantum fluctuations.
}
\label{fig:LiHoF4}
\end{figure}

In 1996 Bitko, Rosenbaum and Aeppli \cite{BRA96} measured the magnetic
properties of LiHoF$_4$ as a function of temperature and a
magnetic field which was applied perpendicular to the preferred
spin orientation. The resulting phase diagram is shown
in Fig. \ref{fig:LiHoF4}.
A minimal microscopic model for the relevant magnetic degrees of
freedom is the Ising model in a transverse field.
Choosing the $z$-axis to be the spin easy axis, the Hamiltonian is given by
\begin{equation}
H = -J \sum_{\langle ij \rangle}S_i^z \, S_j^z - h \sum_i S_i^x~.
\end{equation}
Here $S_i^z$ and $S_i^x$ are the $z$ and $x$ components of the Holmium spin
at lattice site $i$, respectively.
The first term describes the ferromagnetic interaction between the spins,
while the second term is the transverse magnetic field.
For zero field $h=0$, the model reduces to the well-known classical Ising model,
with a ferromagnetic ground state.
With increasing temperature thermal fluctuations of the spins will
reduce the total magnetization, which will eventually vanish at the critical
temperature (about 1.5 K for LiHoF$_4$) above which the system becomes paramagnetic.
The resulting transition is a continuous classical phase transition caused
by thermal fluctuations.

Let us now consider the influence of the transverse magnetic field.
Re-writing $h S_i^x = h/2 (S_i^+ + S_i^-)$, with $S_i^+$ and $S_i^-$
being spin flip operators at site $i$, it is clear that the transverse
field will cause spin flips which are precisely the quantum fluctuations
discussed above.
If the transverse field becomes larger than some critical field
$h_c$ (about 50 kOe in LiHoF$_4$) they will destroy the
ferromagnetic long-range order in the system even at zero temperature.
This transition is a continuous quantum phase transition driven exclusively by
quantum fluctuations, the phase diagram is of the class shown in
Fig.~\ref{fig:qcpvic}b.

For the transverse-field Ising model the quantum--classical mapping
discussed in Sec.~\ref{sec:qucl} can be easily demonstrated at a microscopic
level. Consider a one-dimensional classical Ising chain with the Hamiltonian
\begin{equation}
H_{cl} = -J \sum_{i=1}^N S_i^z \, S_{i+1}^z~.
\end{equation}
Its partition function is given by
\begin{equation}
Z= {\rm Tr}\, e^{-H_{cl}/T} = {\rm Tr}\, \exp(J \sum_{i=1}^N S_i^z S_{i+1}^z / T)
   = {\rm Tr}\, \prod_{i=1}^N M_{i,i+1}~,
\end{equation}
where $\mathbf{M}$ is the so called transfer matrix. It can be represented as
\begin{eqnarray}
\mathbf{M} = \left(
     \begin{array}{cc}
     e^{J/T} & e^{-J/T}\\ e^{-J/T} & e^{J/T}
     \end{array}\right)
     =e^{J/T} (1+ e^{-2J/T} S^x) \approx \exp(J/T + e^{-2J/T} S^x)~.
\nonumber
\end{eqnarray}
Except for a multiplicative constant the partition function of the classical
Ising chain has the same form as that of a single quantum spin in a transverse
field, $H_Q= -h \, S^x$, which can be written as
\begin{equation}
Z= {\rm Tr}\, e^{-H_Q/T_Q} = {\rm Tr}\, e^{h S^x /T_Q} = {\rm Tr} \prod_{i=1}^N
   e^{h S^x /(T_Q N)}~.
\label{eq:quantum Z}
\end{equation}
Thus, a single quantum spin can be mapped onto a classical Ising chain.
Under this mapping, the classical temperature is related to a coupling constant
in the quantum model, $h \equiv e^{-2J/T}$ (being the inverse correlation
length of the classical problem), and the classical system size, $N$, maps onto the inverse
of the quantum temperature, $T_Q$.
These considerations can easily be generalized to a $d$-dimensional transverse-field
(quantum) Ising model which can be mapped onto a $(d\!+\!1)$-dimensional classical
Ising model.
Consequently, the dynamic exponent is $z=1$ here.


\section{Bulk quantum phase transitions}
\label{sec:bulk}

In this section we will review a number of classes of quantum
phase transitions occuring in condensed matter systems.
We will focus our attention on continuous quantum phase transitions,
which are characterized by diverging correlation lengths and times.

The continuous transitions discussed here are so-called bulk transitions,
meaning that degrees of freedom of the whole sample become
critical at the transition point, as is the case with most familiar
phase transitions.
This implies that the singular part of free energy of the
system scales as $L^d$ where $L$ is the linear spatial extension
of the system and $d$ the space dimensionality.
In Sec.~\ref{sec:boundary} we will discuss
so-called boundary transitions where only a subset of the
available degrees of freedom becomes critical.

Most of the following considerations will deal with ``clean'' systems,
where translational invariance is unbroken and quenched disorder is
absent.
The interplay between disorder and criticality is a notoriously
difficult issue -- even in classical systems -- and only understood in some
specific models to date;
we will touch upon this issue in Sec.~\ref{sec:disorder}.


\subsection{Quantum phase transitions and fermions}
\label{sec:fermions}

Quantum critical behavior depends crucially on whether
order parameter fluctuations can couple to low-energy fermionic
excitations (or, more generally, to other non-critical soft modes of
the system).
In the absence of gapless fermions the
order parameter fluctuations are the only low-energy
excitations in the vicinity of the critical point, and the critical
theory can be formulated in terms of the order parameter alone.
Technically, the step of integrating out the fermions from
the action describing the interacting electron system does not lead
to any divergencies.

Critical dynamics can be fundamentally changed if order parameter fluctuations
couple to low-energy fermions, as are present in metals.
Here, integrating out fermions can lead to divergencies.
In some cases, it is believed that a formulation of a critical theory
for the order parameter alone is still possible, however,
the presence of low-energy fermions leads to non-analytic terms in
the effective order parameter theory -- an example is the
Landau damping of critical magnetic fluctuations in metals.

In the following, we specialize the discussion to conventional
order parameters which can be written in terms of local fermion
operators and carry a certain momentum $\bf Q$.
In a translational invariant system, momentum conservation dictates
that an effective coupling between order parameter and fermionic
excitations is only possible if the characteristic wavevector
$\bf Q$ of the order parameter connects Fermi surface points,
as fermions can be scattered off order parameter fluctuations acquiring
a momentum change of $\bf Q$.
Four cases have to be distinguished, characterizing the order parameter
momentum $\bf Q$:
(i) If the order parameter has zero total momentum, as for a ferromagnet,
then it couples to particle-hole pairs at the whole Fermi surface.
We will further discuss this case in Sec.~\ref{sec:metals}.
For an order parameter with finite momentum $\bf Q$, the remaining cases are:
(ii) $\bf Q$ cannot connect points on the Fermi surface.
Then low-energy fermions are scattered into states with higher energy,
perturbation theory in the corresponding coupling is convergent and only gives
a renormalization of parameters.
Thus, the critical behavior is the same as in insulators.
(iii) $\bf Q$ connects a $(d\!-\!2)$-dimensional set of points,
i.e., lines in $d\!=\!3$ or discrete points in $d\!=\!2$.
This leads to damping of the order parameter modes due to the
decay into particle-hole pairs.
At least in $d\!=\!3$ it is believed that this effect can be fully captured
by a Landau damping term in the effective action,
which increases the dynamic critical exponent;
explicit examples are in Sec.~\ref{sec:metals}.
(iv) For so-called perfect nesting, $\bf Q$ can connect
whole parts of the Fermi surface, and then a transition
generically occurs at infinitely small coupling.

An effective coupling between order parameter fluctuations
and low-energy fermions can also be present in unconventional
superconductors which have point or line nodes in their gap
function, because then a Fermi surface still exists, although it
is no longer $(d\!-\!1)$-dimensional.
A particular example is a two-dimensional $d$-wave superconductor,
see Sec.~\ref{sec:sc}.
Similar considerations apply to semimetals.

For unconventional order parameters, e.g., for glassy order or topological order,
the situation is less clear.
It has been proposed that fluctuations of those order parameters can couple
to fermions on the entire Brillouin zone, leading to various scenarios
of ``local criticality''.

In general, a critical point involving both low-energy order parameter
and fermionic modes can be expected to display two time scales,
because order parameter fluctuations can relax both via (damped) order
parameter modes and via fermions.
Consequently, the critical theory should be characterized by
two dynamic critical exponents \cite{rosch}.
A proper RG analysis of such a theory, keeping both low-energy degrees
of freedom, has so far only been carried out for
the ferromagnetic transition \cite{sessions}.


\subsection{Phase transitions with conventional order parameters}

This section describes the physics of continuous quantum phase transitions with
local order parameters, involving, e.g., spin, charge, Peierls, or phase
order.
In space dimensions larger than the classical lower-critical dimension
(see Sec.~\ref{sec:dim}), those transitions are zero-temperature endpoints
of a line of classical phase transitions, Fig.~\ref{fig:qcpvic}b.

\subsubsection{Order parameters.}
\label{sec:OP}

Many order parameters can be expressed as expectation value of combinations
of particle creation and annihilation operators which are {\em local}
in space and time.
The correlation functions of the so-defined local operator ${\cal O}({\bf R}_i,t)$ do
not decay to zero in the long-distance limit in the ordered phase
(long-range order).
As the local quantity can oscillate in space, like the magnetization
in an antiferromagnet, the order parameter $\phi({\bf R}_i)$ is usually
defined as the expectation value of $\cal O$ with the microscopic
oscillations removed,
\begin{equation}
\langle {\cal O}({\bf R}_i) \rangle
= {\rm Re} \left[e^{i {\bf Q} \cdot {\bf R}_i}\phi({\bf R}_i)\right]
\end{equation}
and $\bf Q$ is the ordering wavevector or characteristic momentum.
Sufficiently close to the ordering transition, $\phi({\bf R}_i)$ is now slowly
varying in space, and can be replaced by a continuum field $\phi({\bf r})$.
In a path integral description, we then arrive at a bosonic field $\phi$
carrying momentum $\bf Q$.

The order parameter can have multiple components, i.e., be a spinor, vector, or tensor,
and the underlying microscopics then defines the symmetry properties.
The symmetry in turn determines the allowed terms in the low-energy action and thus
the universality class.
It also decides about the existence of low-energy modes:
an ordered phase which breaks a continuous symmetry supports Goldstone modes.
In addition, the order parameter can carry charge, e.g., if it describes
particle-particle rather than particle-hole pairing;
the presence of long-range Coulomb interactions can then change the nature
of both the collective modes and the transition.

We list a few common examples for order parameters below.
For density waves, the number of components is $N=1$, and the density oscillates
around its average value as
\begin{equation}
\langle \rho ({\bf r}, \tau) \rangle = \rho_{\rm avg} + \mbox{Re} \left[e^{i {\bf Q}_c
\cdot {\bf r}} \phi_c ({\bf r}, \tau)  \right] \,.
\label{chargemod}
\end{equation}
Similar scalar or Ising order parameters also occur, e.g., for alloy ordering
and for Peierls transitions -- in the latter the system undergoes a spontaneous
doubling of the unit cell, induced for instance by strong electron-phonon or
spin-spin interactions.
For spin density waves, we have a vector order parameter, $N=3$ and $\alpha=x,y,z$,
and the spin density modulation is given by
\begin{equation}
\langle S_{\alpha} ({\bf r}, \tau) \rangle = \mbox{Re} \left[e^{i {\bf Q}_s
\cdot {\bf r}} \phi_{s\alpha} ({\bf r}, \tau) \right] \,.
\label{spinmod}
\end{equation}
Here ${\bf Q}_s\!=\!0$ describes a ferromagnet.
In the presence of strong spin anisotropies, fluctuations in some directions
can be frozen out at low energies, then leading to magnetic theories with
$N=1$ or $N=2$.
XY rotor models ($N=2$) can also represent phase variables,
appropriate, e.g., for phase fluctuations in superconductors.

Interestingly, $\phi$ in Eqs. (\ref{chargemod},\ref{spinmod})
is real only for the cases ${\bf Q}=0$ and ${\bf Q}=(\pi,\pi, \ldots)$
(assuming a hypercubic lattice with unit spacing),
otherwise $\phi$ can be complex with the phase describing a sliding degree
of freedom of the density wave;
this phase takes discrete values for commensurate wavevectors.
In the magnetic case, both collinear and non-collinear spin density waves can
be described by Eq. (\ref{spinmod}) \cite{ssrmp}:
\begin{eqnarray}
\mbox{collinear:} &~& \phi_{s\alpha} = e^{i \Theta} n_\alpha ~
\mbox{ with } n_\alpha \mbox{ real} \,, \nonumber \\
\mbox{spiral:} &~& \phi_{s\alpha} = n_{1\alpha} + i n_{2\alpha} ~
\mbox{ with } n_{1,2\alpha} \mbox{ real and } n_{1\alpha} n_{2\alpha} = 0 \,.
\end{eqnarray}
We remind the reader that in spin systems Berry phases are important,
which can render the analysis in terms of a slowly varying order parameter
alone invalid.

Further interesting order parameters are those with an unconventional
symmetry; for fermions bilinears this implies that different parts of
the Fermi surface contribute with different weights or signs.
Examples are superconductors with $p$, $d$, or higher angular momentum
pairing symmetries.
Also particle-hole pairing of unconventional symmetry is possible;
in the charge channel this can be understood as deformation of the Fermi
surface leading to ``nematic'' states -- the transition of a Fermi liquid
into such a state is also known as Pomeranchuk instability.
While most of the listed cases have wavevector ${\bf Q}\!=\!0$, finite $\bf Q$ cases
are possible as well, an example being the so-called $d$-density wave or
staggered flux order \cite{ddw}, with ${\bf Q}=(\pi,\pi)$.


\subsubsection{Insulators.}

The simplest quantum-phase transitions are those where the order parameter
fluctuations do not couple to low-energy fermionic excitations --
this is generically the case in insulators where the gap to charge excitations
is large.

Then, it is sufficient to consider a theory of the order parameter
fluctuations alone.
In most cases, the appropriate continuum theory takes the form
of a $\phi^4$ theory (\ref{phi4}) described in Sec.~\ref{sec:fieldth},
with a dynamic critical exponent $z=1$.
For clean systems, results for static quantities at $T\!=\!0$ can be obtained
from the corresponding classical transition via the quantum--classical mapping.
Most critical properties of such transitions are well understood and
can be found in textbooks \cite{Ma76,Goldenfeld92};
we will not describe them further here.
In contrast, real-time dynamics and transport, in particular at finite temperatures,
cannot be easily extracted from the classical results, and are demanding subjects
of current research \cite{book}.
For disordered systems a number of complications arise, described
in Sec. \ref{sec:disorder}, and only relatively few results are
available.


\subsubsection{Metals: Non-Fermi liquid behavior.}
\label{sec:metals}

Fermi liquids in space dimensions $d\geq 2$ can display ground-state instabilities
towards states with order parameters described in Sec.~\ref{sec:OP}, which
are of considerable practical interest.
In most cases, fermionic low-energy modes couple to the order parameter
fluctuations; many aspects of the resulting theories are not fully
understood to date.
In $d\geq 2$ the quantum critical theories are then above their upper-critical
dimension, therefore non-universal features abound, and microscopic details
like the Fermi surface topology become important.
A unified picture of these transitions is not yet available.

The investigation of quantum phase transitions in Fermi liquids was pioneered
by Hertz \cite{Hertz76} who studied the ferromagnetic transition by a
renormalization group method.
Millis \cite{Millis93} considerably extended this work and computed
finite-temperature crossover functions for several magnetic transitions.
The theoretical approach (commonly called Hertz-Millis theory) proceeds
by integrating out the fermions from the full interacting theory, keeping
only the low-energy order parameter fluctuations in the theory.
Although this is a formally exact step, the coefficients of the resulting
action will in general be non-local in space and time because the fermionic
excitations are gapless.
{\em Assuming} that the action can be expanded in powers and gradients
of the order parameter field $\phi$ and {\em assuming} that it is permissible to
truncate such an expansion, one obtains
\begin{eqnarray}
{\cal S} &=&
\int \frac{d^d k}{(2\pi)^d} T\sum_{\omega_n}
\frac{1}{2} \left[
k^2 + \gamma |\omega_n| + r
\right] |\phi_{\alpha}(k,\omega_n)|^2  \nonumber\\ &&+
\frac{g_0}{4!} \int d^d x d \tau \left( \phi_{\alpha}^2(x,\tau) \right)^2
\label{hertzaf} \,,
\end{eqnarray}
for the case that the ordering wavevector $\bf Q$ connects a $(d\!-\!2)$-dimensional
set of points of the Fermi surface.
Compared to the usual $\phi^4$ theory, the dynamical term is now
replaced by $|\omega_n|$ describing the Landau damping of the order parameter
fluctuations, i.e., the possibility to decay into particle-hole pairs.
The theory (\ref{hertzaf}) has dynamic exponent $z=2$;
for the case of a ferromagnet (${\bf Q}=0$) the damping term gets replaced
by $|\omega_n/k|$, hence $z=3$.
Note that fermionic gap, which occurs in the {\em ordered} phase near the
momenta connected by $\bf Q$, is not captured by (\ref{hertzaf}), therefore
this action is not to be taken seriously at $r<0$, $T=0$ \cite{book}.

The type of quantum critical behavior within Hertz-Millis theory \cite{Millis93}
is dictated by the fact that the interaction term is formally irrelevant above the
upper-critical dimension, therefore only singular {\em corrections} to a Gaussian theory
emerge.
For example, the specific heat of the three-dimensional antiferromagnet
behaves as $C/T = \gamma + A \sqrt{T}$ in the quantum critical region, in other words,
the specific heat coefficient $\gamma$ of the Fermi liquid does {\em not}
diverge upon approaching the quantum phase transition.

Subsequent theoretical work has indicated that the above assumption of a regular gradient
expansion of the action breaks down in a number of important cases.
In the antiferromagnet case, an analysis of the spin-fermion model has indicated
that the Hertz-Millis theory is invalid in $d\!=\!2$ dimensions \cite{chubqcp}.
For the ferromagnetic transition, it has been shown that non-analytic terms
in the spin susceptibility of a usual Fermi liquid lead to corresponding
corrections in the effective order parameter theory \cite{bkv}, which
drastically modify the critical behavior.
Notably, the resulting interaction terms are non-local, and the explicit
form of the four-point vertex is not known, rendering explicit calculations very
difficult in this approach.
Clearly, an improved theory should be {\em local} and should explicitely keep
both the order parameter fluctuations and the low-energy fermionic modes,
making manifest the presence of two time scales in the effective
theory.
Such a calculation has been recently performed for the ferromagnetic
transition in metals.
(In an unconventional superconductor, the situation is somewhat simpler due
to the absence of a full Fermi surface, and both order parameter and fermionic modes
can be kept, see Sec.~\ref{sec:sc}.)
We briefly describe the results for the ferromagnetic transition \cite{sessions}:
In a clean metal with $1<d<3$ the singular behavior of the Fermi liquid
destabilizes any continuous transition with $z=3$, therefore only
a first-order transition to a ferromagnetic state or a $z\!=\!2$ transition to a
state with spiral order appear possible.
In $d=3$ the conclusions are similar, however, the corrections to the $z\!=\!3$
critical point are only logarithmic, and in certain parameter regimes a
continuous transition to a ferromagnetic state may be restored.
In the disordered case, i.e., for diffusive spin dynamics, it was found that
the $z\!=\!4$ fixed point of Hertz is unstable, and instead a new transition with
logarithmic corrections to a Gaussian fixed point with $z=d$ occurs.

On the application side, itinerant ferromagnetic transitions have been found
in MnSi and other intermetallic compounds; in many cases those transitions become
first order at low temperatures, which appears consistent with the theoretical
findings.
Antiferromagnetic transitions appear in both transition-metal and rare-earth compounds.
Great interest has been focussed on the normal-state properties
of the high-temperature superconductors, where simple square-lattice antiferromagnetism
as well as collinear spin and associated charge density waves (``stripes'')
have been detected \cite{tranquada,pnas,castellani,zaanen}.
Those order parameters apparently coexist and/or compete with $d$-wave superconductivity
at low temperatures.
Also, phase transitions involving circulating currents have been
proposed to explain the unusual normal-state properties of high-$T_c$
cuprates~\cite{ddw,varma}.

Much experimental and theoretical work has been performed on heavy-fermion
compounds.
A well-studied compound is CeCu$_{6-x}$Au$_x$, which displays a transition
between a heavy Fermi liquid phase and an antiferromagnetic metallic phase
at $x=0.1$.
Interestingly, at the transition the specific heat coefficient diverges as
$C/T \propto \ln T$, and neutron scattering indicates $\omega/T$ scaling
in the response functions \cite{schroder}.
However, such scaling is only expected below the upper-critical dimension,
see Sec. \ref{sec:dim}, and is clearly at odds with the Hertz-Millis approach.
Some proposals have been made to resolve this inconsistency,
relating the observed quantum critical behavior to the breakdown of Kondo
screening \cite{piers1,edmft,fracfl}.
In Ref. \cite{edmft}, a theory of ``local'' critical behavior has been proposed
in the spirit of the dynamical mean-field approach \cite{dmft};
however, non-trivial behavior emerges here only for two-dimensional systems.
A different approach \cite{fracfl} has associated the suppression of Kondo screening
with the emergence of a non-magnetic ``fractionalized Fermi liquid'' phase,
obtained by suppressing magnetic order in the regime of weak Kondo screening;
see also Sec.~\ref{sec:fract}.


\subsubsection{Superconductors.}
\label{sec:sc}

It is conceivable that a quantum phase transition occurs with
superconductivity being present on both sides including the transition
point.
The main difference to the metallic case discussed above
is that low-energy fermions are gapped due to the presence of the
superconducting gap.
Then, in general the order parameters fluctuations are undamped,
and the phase transitions are in the same universality class as
the ones in insulators.

However, in unconventional superconductors showing gapless points or lines
in momentum space (called nodal points in the following), damping can still occur.
An effective coupling between order parameter fluctuation and fermions is only
present if the wavevector $\bf Q$ of the order parameter fluctuations does
connect two nodal points, see Sec. \ref{sec:fermions}.
Those special cases are ideally treated in a theory which keeps both order
parameter and fermionic excitations, and examples have been considered in
Refs. \cite{bfn,vzs}.
Candidate transitions are those with a particular (fine-tuned) finite
wavevector $\bf Q$, like spin or charge density waves, and
transitions with ${\bf Q}=0$ which do not require any fine tuning.
Examples in the latter class are particle-hole pairing transitions
which lead to ``nematic'' states, and transitions involving the
onset of secondary superconducting pairing \cite{vzs}.
Note that for those transitions the onset of the secondary pairing
occurs at a finite value of the corresponding coupling because
the density of states in the background superconducting state
vanishes at the Fermi level.

We briefly sketch the theory of the transition between two superconducting states
with $d_{x^2-y^2}$ and $d_{x^2-y^2} + i d_{xy}$ pairing symmetry \cite{vzs} -- this is
one of the simplest models where both order parameter fluctuations and
low-energy fermionic excitations can be treated on equal footing.
Importantly, the fluctuations of the secondary $d_{xy}$ order parameter
are represented by an {\em Ising} field ($\phi$), because the fluctuations of its
phase relative to the background $d$-wave pairing are massive.
Thus we simply have
\begin{equation}
{\cal S}_{\phi} = \int \!\! d^2 x d \tau \Bigg[
\frac{1}{2}(\partial_{\tau} \phi)^2 + \frac{c^2}{2} (\nabla_x \phi )^2 +
\frac{r}{2} \, \phi^2 + \frac{g_0}{4!} \, \phi^4 \Bigg];
\label{dsid3}
\end{equation}
as in Eq. (\ref{phi4}).
We explicitely include the low-energy fermions which have a linear spectrum;
for one pair of nodes their action can be written as
\begin{eqnarray}
{\cal S}_{\Psi 1} &=& \int \!\! \frac{d^2 k}{(2 \pi)^2} T \!\sum_{\omega_n}
\Psi_{1a}^{\dagger}  \left(
- i \omega_n + v_F k_x \tau^z + v_{\Delta} k_y \tau^x \right) \Psi_{1a} \,.
\label{dsid1}
\end{eqnarray}
Here $\Psi_{1a} = (f_{1a}, \varepsilon_{ab} f_{3b}^{\dagger})$ is a Nambu spinor
of fermions from opposite nodes (1 and 3) of the $d$-wave superconductor,
$\tau^{\alpha}$ are Pauli matrices which act in the fermionic
particle-hole space, $k_{x,y}$ measure the wavevector from the nodal points and
have been rotated by 45 degrees from $q_{x,y}$ co-ordinates,
and $v_{F}$, $v_{\Delta}$ are velocities.
(An analogous term describes the nodes 2 and 4.)
The full action is then
${\cal S} = {\cal S}_\phi + {\cal S}_{\Psi 1} + {\cal S}_{\Psi 2} + {\cal S}_{\Psi\phi}$
where the final term in the action, $S_{\Psi\phi}$, couples the bosonic and fermionic
degrees of freedom:
\begin{equation}
S_{\Psi\phi} = \int \!\! d^2 x d \tau \Big[ \lambda_0 \phi
\left( \Psi_{1a}^{\dagger} M_1 \Psi_{1a} + \Psi_{2a}^{\dagger} M_2
\Psi_{2a} \right) \Big],
\label{dsid4}
\end{equation}
where $M_{1,2}$ are now specific for the secondary order parameter considered,
for $d_{xy}$ pairing we have $M_1 = \tau^y$, $M_2 = -\tau^y$ \cite{vzs}.
This theory can be analyzed by perturbative RG together with
$\epsilon$ expansion in standard fashion.
The resulting transition has dynamical exponent $z=1$, and the
ordered phase breaks a $Z_2$ symmetry.
However, the transition is not in the Ising universality class due to the
coupling to the fermions.
An explicit calculation of the boson propagator shows that it
contains non-analytic terms in both $k$ and $\omega$ which are different from
simple Landau damping.
The sketched theory has been proposed to describe the strong
damping of nodal fermions in cuprates below the superconducting $T_c$ \cite{vzs}.


\subsection{Quantum phase transitions and disorder}
\label{sec:disorder}

In application to real system the influence of static or quenched disorder
on the properties of a quantum phase transition is an important aspect.
Remarkably, the effect of disorder is not completely understood even for
classical phase transitions.

In a theoretical description, quenched disorder can occur in different ways:
on a microscopic level, e.g., random site energies or bond couplings, or
randomly distributed scattering centers are possible.
In an order parameter field theory, disorder usually translates
into a random mass term for the order parameter fluctuations.
Importantly, applying the quantum--classical mapping (Sec.~\ref{sec:qucl})
to a quantum problem with quenched disorder leads to
a $(d\!+\!z)$-dimensional field theory with strongly anisotropic {\em correlated} disorder
because the disorder is frozen in the time direction.
In some cases lattice effects not captured by the field theory can be
important, this applies, e.g., to all types of percolation problems.
Moreover, disordering a quantum model can lead to random Berry phase terms
which have no classical analogue,
an example are diluted Heisenberg antiferromagnets.

A basis for the discussion of the effect of disorder on continuous transitions
is given by the Harris criterion \cite{Harris74,CCFS86},
which states that if the correlation length exponent $\nu$ of a given
phase transition obeys the inequality $\nu\geq 2/d$, with $d$ the space
dimensionality of the system, then the critical behavior is unaffected
by quenched disorder.
In the opposite case, $\nu < 2/d$, the disorder modifies the critical
behavior, leading either (i) to a new critical
point that has a correlation length exponent $\nu \geq 2/d$ and is thus
stable, or (ii) to an unconventional critical point where the
usual classification in terms of power-law critical exponents looses
its meaning, or (iii) to the destruction of a sharp phase transition.
The first possibility is realized in the conventional theory of random-$T_c$
classical ferromagnets \cite{Grinstein85}, and the second one is probably
realized in classical ferromagnets in a random
field \cite{Fisher86}.
The third one has occasionally been attributed to
the exactly solved McCoy-Wu model \cite{McCoyWu68}.
This is misleading,
however, as has been emphasized in Ref.\ \cite{Fisher95};
there is a sharp, albeit unorthodox, transition in that
model, and it thus belongs to category (ii).

For first-order transitions, Imry and Wortis \cite{ImryWortis}
have proposed a criterion
stating that below a certain critical dimension (which is usually 2)
the transition will be significantly rounded due to domain formation.
The question of whether the possibly resulting continuous transition
shows universal scaling is not settled \cite{Cardy98}.

Independent of the question of if and how the critical behavior is affected,
disorder leads to very interesting phenomena as a phase transition is
approached.
Disorder in classical systems generically decreases the critical temperature $T_c$
from its clean value $T_c^0$. In the temperature region $T_c<T<T_c^0$ the system
does not display global order, but in an infinite system one will find
arbitrarily large regions that are devoid of impurities, and hence show
local order, with a small but non-zero probability that usually decreases
exponentially with the size of the region.
These static disorder fluctuations
are known as ``rare regions'', and the order parameter fluctuations induced
by them as ``local moments'' or ``instantons''. Since they are weakly coupled,
and flipping them requires to change the order parameter in a whole
region, the local moments have very slow dynamics.
Griffiths \cite{Griffiths69} was the first to show that they lead to a
non-analytic free energy everywhere in the region $T_c<T<T_c^0$, which
is known as the Griffiths phase, or, more appropriately, the Griffiths
region. In generic classical systems this is a weak effect, since the
singularity in the free energy is only an essential one.
(An important exception is the McCoy-Wu model \cite{McCoyWu68}.)

Turning now to quantum phase transitions, it is expected that disorder
has a stronger effect compared to the classical case due to the disorder
correlations in time direction.
Consequently, the Harris criterion is still given by $\nu < 2/d$ (not $d+z$).
A prototypical and well-studied model is the quantum Ising chain in a
transverse random field, investigated by Fisher \cite{Fisher95} and others,
where the phase transition is controlled by a so-called infinite-randomness
fixed point, characterized by a wide distribution of couplings and
activated rather than power-law critical behavior.
The Griffiths singularities are also enhanced compared to the classical
case: a number of observables display power-law singularities with
continuously varying exponents over a finite parameter region in the disordered
phase.
Numerical simulations \cite{rs2d} suggest that random-singlet phases
and activated criticality may not be restricted to one-dimensional systems,
raising the possibility that exotic critical behavior dominated by rare regions
may be common to certain quenched-disorder quantum systems, in particular those
with Ising symmetry.
Recent investigations of two-dimensional diluted bilayer antiferromagnets
with Heisenberg symmetry indicate conventional critical behavior, and
in addition an interesting interplay of quantum and geometric criticality at the
percolation threshold \cite{SandvikQP}.

All the examples mentioned so far display undamped order parameter dynamics
($z=1$ in the clean system).
In contrast, in itinerant electron systems the dynamics is overdamped due
the coupling to fermions
($z>1$ in the clean limit, see Secs.~\ref{sec:fermions} and \ref{sec:metals}),
which also changes the behavior upon introduction of disorder.
It has been shown recently that for Ising order parameter symmetry and
overdamped dynamics a continuous transition is always rounded due to
the presence of ordered islands arising from a $1/r^2$ interaction in the
effective classical model \cite{TomDisIsing}.
For continuous symmetries it is likely that a sharp transition survives.
A prototype example is the disordered itinerant antiferromagnet, where
it has been shown that a finite-disorder fixed point is destabilized by
the effects of rare regions. The RG indicates run-away flow to large disorder,
and definite conclusions regarding the transition cannot be drawn \cite{bkrs}.
In the context of heavy-fermion compounds, it has been suggested that
non-Fermi liquid behavior can occur in the quantum Griffiths region of a
disordered Kondo lattice model \cite{antonio}.

Systems which are {\em dominated} by disorder and frustration can have a variety of new
stable phases not known from clean systems, e.g., spin or charge glasses and
various types of infinite-randomness phases.
Associated phase transitions then belong to new universality classes,
an interesting example being the quantum spin glass transition, i.e.,
the $T\!=\!0$ transition between a paramagnet and a spin glass \cite{book}.
Such transitions may be relevant for certain strongly disordered heavy-fermion
compounds, but also speculations in the context of high-$T_c$ superconductivity
have been put forward \cite{qcpopt}.


\subsection{Metal--insulator transitions}
\label{sec:mit}

Metal--insulator transitions are a particularly fascinating and only
incompletely understood class of quantum phase transitions.
Conceptually, one distinguishes between transitions in models of
noninteracting electrons, arising from solely from lattice effects,
and transitions of interacting electrons.
Examples in the first class are band or Peierls metal--insulator transitions,
as well as the disorder-driven Anderson transition.
The prominent example in the second class is the Mott transition
of clean interacting electrons.
At the Anderson transition, the electronic charge diffusivity $D$ is
driven to zero by quenched disorder, while the thermodynamic
properties do not show critical behavior.
In contrast, at the Mott transition the thermodynamic density susceptibility
$\partial n/\partial\mu$
vanishes due to electron-electron interaction effects. In either case, the
conductivity $\sigma = (\partial n/\partial\mu)D$ vanishes at the
metal--insulator transition.
It is worth emphasizing that a sharp distinction between metal and insulator
is possible only at $T\!=\!0$;
in some cases, a {\em first-order} finite-temperature transition
between a ``good'' and a ``bad'' metal can occur.

Let us briefly discuss the disorder-driven metal--insulator transition.
Introducing quenched disorder into a metallic system, e.g., by adding
impurity atoms, can change the nature of the electronic states from spatially
extended to localized \cite{Anderson58}.
This localization transition of disordered {\em non-interacting} electrons,
the Anderson transition, is comparatively well
understood, see Ref.~\cite{KramerMacKinnon93}.
The scaling theory of localization \cite{AALR79} is based on the assumption of
hyperscaling and predicts that
in the absence of spin-orbit coupling or magnetic fields all states
are localized in one and two space dimensions for arbitrarily weak
disorder.
Thus, no true metallic phase exists for $d=1,2$.
In contrast, in three dimensions there is a phase transition from
extended states for weak disorder to localized states for strong disorder.
These results of the scaling theory are in agreement with large-scale
computer simulations of non-interacting disordered electrons.
The effect of weak disorder can be captured by perturbation theory,
leading to the well-known ``weak-localization'' quantum corrections to the
conductivity, which are divergent in $d=1,2$.
A field-theoretic description of the Anderson transition
has been pioneered by Wegner \cite{wegner}.
Starting from electrons in a random potential, the disorder average is performed
using the Replica trick, and the arising interaction terms can be decoupled
using complex matrix fields.
The theory then takes the form of
a non-linear sigma model in terms of these matrix fields; it has been
subsequently analyzed at the saddle-point level and by RG methods,
with results consistent with the scaling theory \cite{bkmott}.

A different limiting case is the interaction-driven Mott transition
of clean electrons which can occur at commensurate band fillings.
The most studied model showing such a transition is the one-band
Hubbard model, with the Hamiltonian
\begin{equation}
H\, =\, - \sum_{\langle ij\rangle \sigma} t_{ij}
      (c^\dagger_{i\sigma} c^\pd_{j\sigma} +
       c^\dagger_{j\sigma} c^\pd_{i\sigma})
    + U \sum_i  n_{i\uparrow} n_{i\downarrow}
\,,
\label{H_HUBBARD}
\end{equation}
defined on a regular lattice, with hopping energies $t_{ij}$ and
an on-site Coulomb repulsion $U$.
At half filling, the system is metallic at $U=0$, but insulating at $U\to\infty$
because large $U$ prevents doubly occupied sites and thus completely suppresses
hopping.
In general, an interaction-driven metal--insulator transition may
be accompanied by the occurrence of magnetic order (Mott-Heisenberg transition).
In contrast, the transition from a paramagnetic metal to a {\em paramagnetic}
insulator is termed Mott-Hubbard transition.
Non-perturbative techniques are necesssary to analyze the Mott transition because
it generically occurs at intermediate coupling (except for perfect nesting
situations).

The Hubbard model (\ref{H_HUBBARD}),
both at and away from half-filling, has been extensively
studied in the context of high-temperature superconductivity, and
a plethora of phases have been proposed as function of parameters and
doping.
However, many properties of this apparently simple one-band model in two and three
dimensions are still under debate, as analytical calculations often have
to rely on uncontrolled approximations, and numerical methods are limited
to very small system sizes.
In one dimension, special techniques have provided a number of exact answers,
see Ref.~\cite{florian}.
Beyond that, considerable progress has been made in the framework of dynamical
mean-field theory (DMFT) \cite{dmft}, which represents a local approximation to the
self-energy and becomes exact in the limit of infinite space dimensions.
In this approach, the lattice model is mapped onto a quantum impurity model
supplemented by a self-consistency condition.
Extensive numerical studies have established that the $T=0$ Mott transition
in the DMFT limit is characterized by the {\em continuous} vanishing of
the quasiparticle weight upon increasing $U/t$;
however, the insulator displays a preformed gap, and no dynamic
critical behavior is present \cite{dmft,dmftmott}.

In reality, both Coulomb interaction and disorder are present, and their
interplay is poorly understood.
The conventional approach to the problem of disordered {\em interacting}
electrons is based on a perturbative treatment of both disorder and
interactions \cite{AltshulerAronov85,LeeRamakrishnan85}.
It leads to a scaling theory and a related field-theoretic formulation
of the problem \cite{Finkelstein83}, which is a generalization of Wegner's
theory for non-interacting electrons \cite{wegner}, and was
later investigated in great detail within the framework of the renormalization
group (for a review see Ref.~\cite{bkmott}).
One of the main results is that in the absence of external symmetry breaking
(spin-orbit coupling or magnetic impurities, or a magnetic field)
a phase transition between a {\em normal} metal
and an insulator only exists in dimensions larger than two, as was the case for
non-interacting electrons. In two dimensions the results of this approach
are inconclusive since the renormalization group displays runaway flow to
zero disorder but infinite interactions.
Furthermore, it has not been
investigated so far, whether effects of rare regions (Sec.~\ref{sec:disorder})
would change the above conclusions about the metal--insulator
transition.
It is known that local moments tend to form in the vicinity
of the transition, however, their role in the critical properties has
not been clarified.
Conceptually, it is clear that a perturbative approach is not applicable
in a regime of strong interactions.

Experimental work on the disorder-driven metal--insulator transition,
mostly on doped semiconductors, carried out before 1994 essentially
confirmed the existence of a transition in three dimensions \cite{hvlsip}
while no transition was found in two-dimensional systems.
Therefore it came as a surprise when experiments
on Si-MOSFETs \cite{KMBFPD95} revealed indications of a
true metal--insulator transition
in two dimensions.
These experiments have been performed on high-mobility
samples with a low carrier density.
Therefore, electronic interactions are strong: for an electron density of
$10^{11}{\rm cm}^{-2}$ the typical Coulomb energy is about 10 meV
while the Fermi energy is only about 0.5 meV.
Interaction effects are a likely reason for this new
behavior in two dimensions, however, a complete understanding
has not yet been obtained.
It is not settled whether the observations are associated
with a true zero-temperature transition, or whether they
represent a intermediate-temperature crossover phenomenon.
Recent experiments indicate glassy charge dynamics on both the metallic and
the insulating side of the apparent transition,
further pointing to an interesting interplay between disorder and interactions.

A number of ideas developed in context of the Anderson transition also apply to
transitions in quantum Hall systems.
In each quantum Hall plateau phase, electrons are localized due to disorder,
and the longitudinal $T=0$ conductivity vanishes.
Transitions between different quantum Hall states involve a diverging correlation
length and scaling behavior of the conductivity.
In addition to scaling theories, approaches based on quantum tunneling network models,
as first proposed by Chalker and Coddington, have been successful in explaining
a number of experimental findings.
However, at present the role of electron interactions at the quantum Hall
critical points has not been thoroughly clarified.
For a full exposure we refer to recent review articles \cite{sondhi,huckestein}.

Summarizing, metal--insulator transitions, in particular those occurring at
intermediate or strong interactions, are in the focus of current research
activity, and many aspects, including the interplay with lattice effects
and magnetism, remain to be understood.


\subsection{Superfluid--insulator transitions}
\label{sec:sit}

Transitions between superfluid and insulating states can, like metal--insulator
transitions, be driven by disorder or by strong Coulomb interaction;
in the case of a charged superfluid, i.e., a superconductor, applying a magnetic field
can also lead to a superconductor--insulator transition.
Both the disorder and field-driven transitions have been extensively studied
in the context of thin superconducting films.
Interaction-driven (Mott) transitions can be realized in Josephson junction arrays where
the ratio between the inter-island Josephson coupling and the intra-island Coulomb
interaction can be varied \cite{sondhi}.
Very recently, clear signatures of a Mott superfluid--insulator transition
have been observed in a system of ultracold atoms in an optical
lattice~\cite{bloch}.

Usually, models of bosons are employed to describe the superfluid.
For the disorder-free case with interactions one is lead to the boson Hubbard model,
introduced in Sec.~\ref{sec:fieldth},
which contains a tight-binding hopping energy $t$
and an on-site repulsion $U$.
The ground-state phase diagram is well understood:
In a plane parametrized by the chemical potential $\mu/t$ and the
ratio $t/U$ the large-$t$ regime is superfluid whereas in the
small-$t$ regime Mott phases occur where the density is fixed to integer
values.
The transitions are described by an O(2) rotor model capturing the condensate phase
dynamics provided the density is fixed -- this is the case on the tips of the
so-called Mott lobes of the phase diagram.
Otherwise, the complex condensate density itself is the appropriate order
parameter, and the transition is in the dilute Bose gas class \cite{book}.

In the presence of disorder, the situation is somewhat more complicated.
In addition to the Mott insulator and superfluid phases a
Bose glass phase appears, which is believed to be insulating.
Particular attention has been focussed on the two-dimensional case
where the scaling dimension of the conductivity turns out to be zero.
Thus, scaling theories predict a universal metallic resistance exactly
at the transition point between superconductor and insulator \cite{sondhi,z1}.
An important feature of the theory is the relevance of long-range
Coulomb interaction at the transition, which leads to a dynamical critical
exponent being unity \cite{z1}.
A duality description of the two phases, involving bosons and vortices,
has been put forward:
in the superconducting phase the bosons are condensed and the vortices
localized whereas in the insulating (Bose glass) phase the
bosons are localized and the vortices condensed -- this corresponds
to a superfluid of vortices \cite{SIT}.

While these theories were confirmed by a number of experimental studies,
they have been called into question by recent experiments on superconducting films
in a field, which suggest the existence of a finite intermediate regime of metallic
behavior.
A number of proposal have been put forward, ranging from a metallic
Bose glass, or a true Bose metal, to metallicity produced by the influence of
dissipation \cite{mason}.
Other speculations concern the possible influence of {\em fermionic}
excitations, which are usually neglected in boson-only models.
In this context one has to keep in mind that in two dimensions a fermionic
metallic phase is believed to be absent due to weak localization;
however, the experimental findings on two-dimensional electron gases mentioned in
Sec.~\ref{sec:mit} have called this into question as well.
Thus, the fascinating fields of metal--insulator and superconductor--insulator
transitions are firmly connected, and the interplay between disorder and
interactions remains an important topic of future research.


\subsection{Phase transitions involving topological order}
\label{sec:topo}

Phase transitions can involve the onset of so-called topological order;
in this case the ordered state is not characterized by a local order
parameter; however, the two phases can be distinguished from their global
properties.
In many cases, topological order can be rephrased as the suppression of
topological defects, either in physical (e.g. spin) or emergent (gauge field)
degrees of freedom.
The transition from a topologically ordered to a disordered state
can be described as the proliferation of topological defects.
Topological order occurs in different contexts, e.g., in the quantum
Hall effect, in superconductors, and in spin systems;
we will describe a few interesting transitions below.

\subsubsection{Kosterlitz-Thouless transition.}
\label{sec:KT}

A well-known classical example is the finite-temperature transition
of the $d\!=\!2$ XY model, i.e., a model where spins are confined to the $x$-$y$ plane
which is equivalent to a O(2) rotor model.
At high temperatures the spins are disordered with exponentially
decaying correlations, whereas at low, non-zero temperatures the spin
display quasi-long-range order with power-law correlations.
This quasi-ordered phase has topological vortex excitations
which are suppressed at low temperatures, i.e., the density of
vortex-antivortex pairs is small.
The two phases are separated by a Kosterlitz-Thouless
transition at $T_{\rm KT}$ where vortex pairs unbind upon increasing temperature;
the transition is accompanied by a universal jump in the superfluid density.
The XY model is of relevance for the phase dynamics of two-dimensional
superconductors and has been discussed extensively in the context
of underdoped high-temperature superconducting cuprates.

A Kosterlitz-Thouless {\em quantum} phase transition occurs at $T\!=\!0$
in the $(1\!+\!1)$-dimensional XY model, or equivalently in the
O(2) rotor model in $d\!=\!1$ --  this follows straightforwardly
from the quantum--classical mapping of Sec.~\ref{sec:qucl}.
Technically, the physics is contained in a model of the
sine-Gordon type, where the action of a free field describing the phase dynamics
is supplemented by a term arising from vortex tunneling events \cite{book}.

The model can be analyzed by conventional RG methods;
interestingly, formally the same RG equations occur in the context
of the Kondo model and will be discussed in Sec.~\ref{sec:kondo}.
The RG analysis shows that the Kosterlitz-Thouless transition
does not correspond to a critical, i.e., unstable fixed point in the RG sense,
therefore ``conventional'' quantum critical behavior (as in Fig.~\ref{fig:qcpvic})
and the corresponding power laws do {\em not} occur near in the vicinity of
the transition point.
Connected to that, the Kosterlitz-Thouless transition does not yield a
singularity in any derivative of the thermodynamic potential at the transition,
therefore it is sometimes called infinite-order transition.

\subsubsection{Fractionalization transitions.}
\label{sec:fract}

In the context of strongly correlated electron systems in dimensions $d>1$,
phases with ``fractionalized'' elementary excitations,
carrying quantum numbers different from multiples of those of the electron,
have attracted considerable attention in recent years.
Fractionalization is usually discussed by re-casting the original model
into a theory of new (fractionalized) quasiparticles, e.g., spinons and chargons,
interacting with each other through a gauge field:
the quasiparticles will carry a gauge charge.
A fractionalized phase corresponds to a deconfined phase of the gauge theory,
i.e., the effect of the gauge field is weak.
In contrast, if the gauge field is confining, the fractionalized quasiparticles
are bound at low energies, resulting in excitations with conventional quantum
numbers.
The fractionalization transition between such two phases corresponds to a
confinement--deconfinement transition of a certain gauge theory.
The transition can be either driven by the condensation of
topological defects of the gauge field, or by the condensation
of matter (Higgs) fields which are coupled to the gauge field.

The most prominent examples of gauge theories in the context of
fractionalization are the ones with $Z_2$~\cite{z2a,z2b} and U(1)~\cite{bosfrc}
symmetry of the gauge field.
Based on early work on the phase diagram of the U(1) gauge theory
by Fradkin and Shenker \cite{fradsh}, it is believed that
$Z_2$ fractionalization can occur in $d= 2,3$ whereas
U(1) fractionalization can only occur in $d=3$.
The deconfined phases of those theories possess topological order
which is associated with the suppression of topological defects
in the gauge field;
this order manifests itself also in a topological ground state degeneracy
if the system is subjected to periodic boundary conditions.

Fractionalization has been intensively discussed in the context
of low-dimensional undoped and doped Mott insulators,
with particular focus on high-$T_c$ superconductivity.
In fact, Anderson's resonating valence bond state for two-dimensional
antiferromagnets \cite{rvb} is one of the first
proposals for a fractionalized spin liquid.
A powerful description of such a singlet liquid state is provided
by $Z_2$ gauge theory models \cite{z2a,z2b}, see
Ref.~\cite{ssrmp} for further discussion.
Experimental tests for fractionalization, based on the existence of vortices
(so-called visons) in the $Z_2$ gauge field, have been undertaken,
but no topological order has been detected so far in the cuprates.

Significant progress has been made in constructing explicit models
which show fractionalization; these involve simple models
of bosons and fermions on regular lattices \cite{bosfrc}.
A notable achievement is the demonstration of fractionalization
in the triangular lattice quantum dimer model \cite{moessner};
furthermore indications for fractionalization have been found
in numerical investigations of ring-exchange models on the triangular
lattice, and charge fractionalization has been demonstrated in
a correlated model on the pyrochlore lattice~\cite{fulde}.

Recently, proposals for fractionalized phases in Kondo-lattice models
have been put forward \cite{fracfl}.
Here, in the presence of strong quantum fluctuations the local moments can form
a fractionalized spin liquid in a regime where the inter-moment interaction
dominates over Kondo screening.
The topological order protects this spin liquid against small perturbations,
thus the conduction electrons will be essentially decoupled from the
local moments.
This peculiar paramagnetic phase has been termed ``fractionalized Fermi liquid'',
as it displays both Fermi-liquid-like particle-hole excitations and fractionalized
spinon excitations, arising from the conduction electron and local moment
subsystems, respectively.
Increasing the Kondo coupling eventually drives the system through a (Higgs)
confinement transition into the usual heavy Fermi liquid phase.
Approaching from the Fermi liquid side, this zero-temperature transition can be
understood as the breakdown of Kondo screening.
The quantum critical region of this fractionalization transition displays novel
critical behavior \cite{fracfl} which may be related to experimental findings in heavy-fermion
systems like CeCu$_{6-x}$Au$_x$ \cite{schroder}.

It is clear that fractionalization in $d>1$ is a fascinating field which
can lead to a plethora of new phenomena in condensed matter physics.
Although experimental evidence for the occurrence of such phases
is lacking to date, we have to keep in mind that the experimental distinction can
be very subtle due to the absence of a conventional order parameter;
it could very well be the case that fractionalization is realized in
recently studied strongly correlated materials.
Therefore, further theoretical work predicting properties and proposing
stringent tests of fractionalization is called for.


\subsection{Competing orders}
\label{sec:compete}

The physics of competing orders has been a central theme in
low-dimensional correlated electron systems over the last years.
In particular, in two dimensions extensive studies of microscopic models
have shown that a variety of phases compete, and the actual ground state
may depend sensitively on details of the model parameters.
The interplay of different orders appears crucial for the
understanding of the complex phase diagrams of various materials,
with high-temperature superconductors being a paradigmatic example.

On the theoretical side, competing orders imply the existence of multiple
order parameters.
A simple phenomenological description involves a Landau-type theory of coupled
order parameter fields, in which the form of the allowed terms follows
from symmetry arguments.
Let us discuss the case of two order parameters $\phi_A$ and $\phi_B$, which
break {\em different} symmetries of the Hamiltonian:
then the simplest allowed coupling is usually of the density-density type
$|\phi_A|^2 |\phi_B|^2$.
The Landau theory will generically permit four phases: disordered, pure A, pure B,
and coexistence of A and B.
Tuning the system between the pure A and pure B phases leads either to
{\em two} continuous transitions, with a coexistence phase or disordered phase
in between, or to a first-order transition between A and B;
the occurence of a {\em single} continuous transition requires
additional fine tuning of parameters in the action.

In the context of high-$T_c$ superconductors, the competition between magnetism
and superconductivity has been studied extensively.
On the one hand, static magnetism tends to suppress superconductivity,
on the other hand, magnetic fluctuations are believed to mediate pairing
(perhaps supplemented by phonons).
The intimate relation between antiferromagnetism and $d$-wave superconductivity
has been suggested to arise from an (approximate) higher symmetry of
the underlying model, namely an SO(5) symmetry unifying the two-component
superconducting and the three-component magnetic order
parameter~\cite{sczhang}.
Other proposals for orders competing and/or coexisting with superconductivity
involve charge-stripe order \cite{castellani,zaanen},
nematic order \cite{nematics},
circulating currents \cite{ddw,varma},
and more exotic orders related, e.g., to spin-charge separation~\cite{z2b}.

\begin{figure}[!t]
\epsfxsize=5.5cm
\epsffile{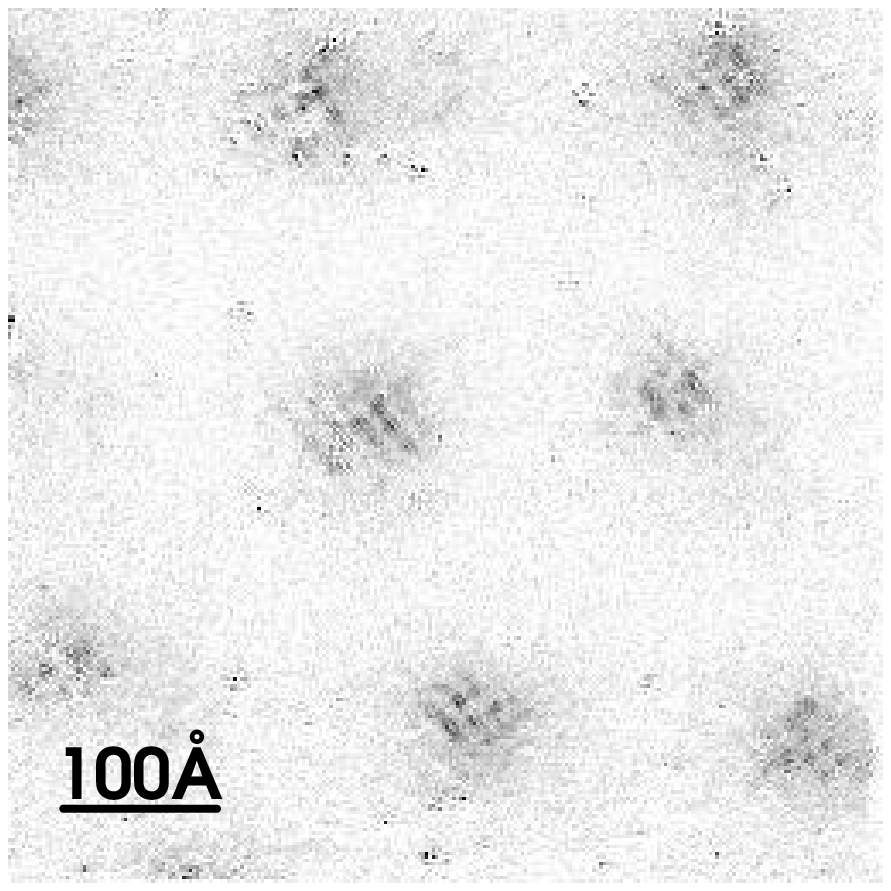}
\vspace*{-5.3cm}
\hspace*{6.7cm}
\epsfxsize=8.7cm
\epsffile{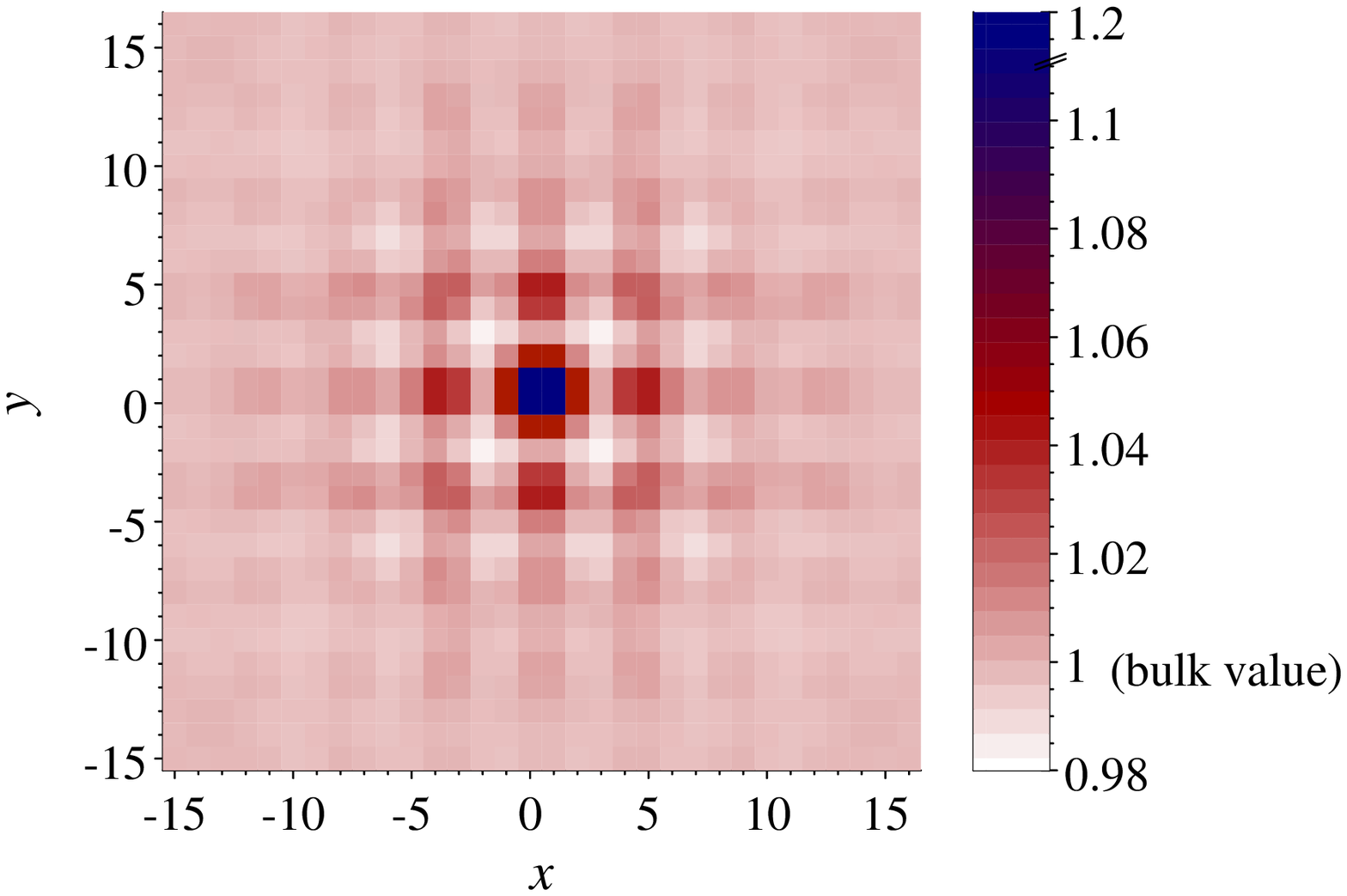}
\caption{
Left:
Grayscale plot of the local DOS as measured by STM
at the surface of Bi$_2$Sr$_2$CaCu$_2$O$_{8+x}$ in a magnetic field of 7 T,
integrated over an interval of subgap energies
(after \protect\cite{fieldstm}).
The dark regions can be identified with the location of vortices.
Around each vortex a checkerboard modulation is clearly visible;
with the lattice constant being about 3.9 \AA\ for Bi$_2$Sr$_2$CaCu$_2$O$_{8+x}$,
the modulation period is found to be four lattice spacings.
Right:
Theoretical computation of the local DOS within a pinning model for the dynamic spin density
wave fluctuations \protect\cite{pin} -- here the unit of length is the lattice constant.
(The crystal axes are rotated by approximately 45 degrees in the left plot.)
}
\label{fig:pin}
\end{figure}

Recent neutron scattering and scanning tunneling microscopy (STM) experiments
indicate that important competitors of superconductivity
are incommensurate spin and associated charge density waves (CDW).
The spin density waves (SDW) correspond to a collinear spin order at
wavevectors ${\bf Q}_{sx}=(\pi\pm\epsilon,\pi)$ and ${\bf Q}_{sy}=(\pi,\pi\pm\epsilon)$
where $\epsilon$ depends on doping.
On general symmetry grounds, such spin order ($\phi_{s\alpha}$) is accompanied
by charge order ($\phi_c$) at wavevectors
${\bf Q_c} = 2 {\bf Q}_s$; formally this follows from the
existence of a {\em cubic} term $(\phi_{s\alpha}^2 \phi_c^\ast + c.c.)$
in the field theory, which is allowed only if ${\bf Q}_c = 2 {\bf Q}_s$ due to
momentum conservation.
(Note that CDW order can exist without simultaneous SDW order.)
Incommensurate spin and charge orders, coexisting with superconductivity
at low $T$, have been detected in Nd and Eu-doped La$_{2-x}$Sr$_x$CuO$_4$ \cite{tranquada}.
A number of other experimental results \cite{lake,fieldstm}
can also be explained by proximity to a quantum critical point at which
spin/charge order disappears:
even in optimally doped compounds which show no long-range spin order
a suitable description of the collective spin and charge excitations
is given by a theory assuming the vicinity of an ordering transition at
lower doping.
Then, the additional order parameters are fluctuating \cite{ssrmp,kivrmp},
and will show up in suitable dynamic measurements.
Charge fluctuations can also become static due to impurity pinning,
and thus be detectable, e.g., by STM.
An applied magnetic field can enhance competing spin/charge fluctuations or even
stabilize static SDW/CDW order by suppressing the competing superconductivity
due to vortex superflow \cite{demler,ssrmp}.
Recent STM experiments on optimally doped Bi$_2$Sr$_2$CaCu$_2$O$_{8+x}$ in a
magnetic field show that a checkerboard modulation occurs in the local
density of states near vortex cores \cite{fieldstm}, see Fig.~\ref{fig:pin}.
This supports the idea of competing orders, and a likely explanation is
that dynamic collective charge fluctuations are enhanced in the regions
of weakened superconductivity and pinned by local imperfections, like
the vortex cores themselves \cite{pin}.
A detailed modelling of the energy dependence of the STM signal indicates that it is
best described by a checkerboard modulation in microscopic bond rather than site
variables \cite{vj2002}.

\begin{figure}[t]
\epsfxsize=9cm
\centerline{\epsffile{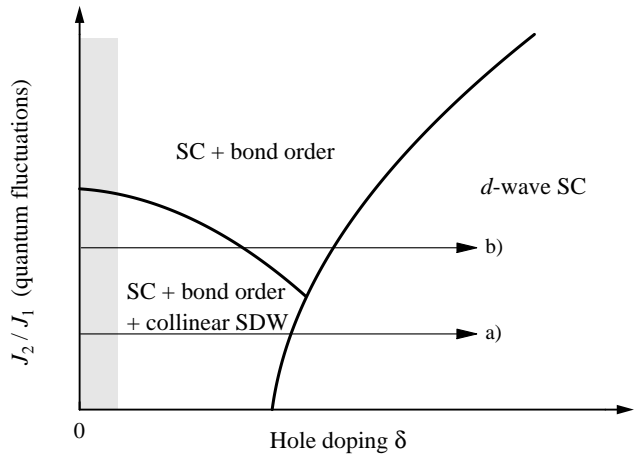}}
\caption{
Schematic $T\!=\!0$ phase diagram (after
Refs.~\protect\cite{vzs,ssrmp}) for the high temperature
superconductors.
The vertical axis denotes a parameter which can destroy
antiferromagnetic order in the undoped limit, like
the ratio of the near-neighbor exchange interactions;
the horizontal axis is the hole concentration, $\delta$.
The ground state is insulating at small doping due to Coulomb
interactions (shaded).
All other phases at finite $\delta$ show superconductivity
which can coexist with bond/charge order and
possibly collinear magnetic order at lower doping.
The two arrows a) and b) correspond to the possible hole doping
evolution of different compounds.
}
\label{fig:phase}
\end{figure}

At the microscopic level, the most important aspect of cuprates is
that superconductivity is obtained from doping a Mott insulator.
The generic phase diagram of doped Mott insulators is not understood
to its full extent, however, it is clear that various spin and charge
ordered phases, superconductivity, and more exotic phases could be
realized.
One simple scenario, based on large-$N$ calculations for an extended $t-J$ model,
has been put forward in Ref.~\cite{vzs}:
starting from a paramagnetic Mott insulator with bond order, doping
naturally leads to superconducting phases coexisting with bond order, i.e.,
stripe-like charge density modulations.
The resulting phase diagram is reproduced in Fig.~\ref{fig:phase}.
The interplay between superconductivity and bond order has been
studied in some detail in Refs.~\cite{vzs,vj2002}:
The ordering wavevector shows a characteristic evolution with doping,
and period-4 structures have been found stable over a significant
doping and parameter range.
Phase diagrams with related physical ingredients, but some significant differences,
appear in Refs. \cite{zaanen,nematics}.
We note that a number of experimental results indicate spin freezing
into a glassy state below optimal doping, as has been pointed out by
Panagopoulos {\em et al.} \cite{qcpopt} -- this may be expected in
the presence of disorder in a region of the phase diagram where spin ordering
occurs in the clean limit.

Besides the high-$T_c$ compounds, numerous other strongly correlated materials
display competing orders.
Evidence for co-existing antiferromagnetism and superconductivity has also been
found in heavy-fermion compounds,
e.g., CePd$_2$Si$_2$, CeCu$_2$Si$_2$, and CeIn$_3$.
However, in these materials the nature of the superconducting state
has not been established to date.
Other examples are colossal magnetoresistance manganites,
ruthenium oxides, and organic conductors of the TMTSF and BEDT-TTF type.


\section{Boundary quantum phase transitions}
\label{sec:boundary}

This section is devoted to the special class of boundary phase transitions.
Such transitions occur in systems which can be divided into a bulk part
with space dimension $d$ and another part with dimension $d_b < d$ --
this ``boundary'' part can be either a surface or interface, or even
a single impurity ($d_b=0$) embedded into the bulk.
At a boundary transition only the boundary degrees of freedom
undergo a non-analytic change.
At a continuous boundary phase transition, there is a singular part
in the free energy which scales as $L^{d_b}$ where $L$ is the linear
dimension of the system.

Boundary transitions have been extensively discussed in the context of surface
transitions in magnets~\cite{boundarypt}, where under certain conditions the
surface spins can order at a temperature above the bulk ordering
temperature.
The fascinating aspects of quantum criticality can also be
found in boundary transitions.
Here, we shall describe a few zero-temperature transitions in quantum impurity
systems, which have been of much interest in diverse fields
such as unconventional superconductors, heavy fermions, quantum dot systems,
and quantum computing.


\subsection{Kondo effect in metals and pseudogap Fermi systems}
\label{sec:kondo}

\begin{figure}[!t]
\epsfxsize=10cm
\centerline{\epsffile{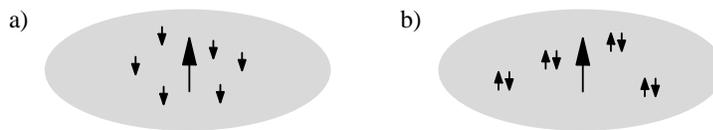}}
\caption
{
a)
Visualization of Kondo screening in metals:
an impurity spin is surrounded by conduction electrons with primarily opposite spin.
b)
Kondo screening in superconductors is suppressed because the
low-energy electrons are locked into Cooper pairs.
However, for $J_K$ larger than the Cooper pair binding energy
screening is again possible -- this can be interpreted as Bogoliubov quasiparticles
bound locally to the impurity spin.
}
\label{fig:kondovis}
\end{figure}

A magnetic impurity spin $\bf S$ embedded in a bulk system of non-interacting
conduction electrons (Fig.~\ref{fig:kondovis}a) can be
described by the so-called Kondo Hamiltonian,
\begin{equation}
   H = \sum_{k\sigma} \varepsilon_k c^\dagger_{k\sigma}
        c_{k\sigma} + J_K {\bf S} \cdot {\bf s}(0) \ ,
\end{equation}
where ${\bf s}(0) = \sum_{\bf kk'\sigma\sigma'} c^\dagger_{{\bf k}\sigma} {\bf \sigma}_{\sigma\sigma'}
c_{{\bf k '}\sigma'}$
is the conduction electron spin at the impurity site ${\bf r}\!=\!0$.
The Kondo effect in metals, occuring at any positive value of the Kondo coupling $J_K$,
is by now a well studied phenomenon in many-body physics.
In essence, at low temperatures the conduction electrons
and the impurity spin form a collective many-body state
with zero total spin.
The scattering of the conduction electrons off the impurity gives rise to
non-trivial transport behavior, namely a temperature minimum in the electric
resistivity of a metal with dilute magnetic impurities.
For small $J_K>0$, the low-energy physics of an isolated impurity
is completely determined by a single energy scale, $k_B T_K$,
where $T_K$ is the Kondo temperature.
The impurity spin is fully screened by the conduction electrons
in the low-temperature limit, $T \ll T_K$.
$T_K$ depends exponentially on the density of states (DOS) at the
Fermi level, $\rho_0 = \rho(\epsilon_F)$, and the Kondo coupling $J_K$ \cite{hewson}.

\begin{figure}[!t]
\epsfxsize=12cm
\centerline{\epsffile{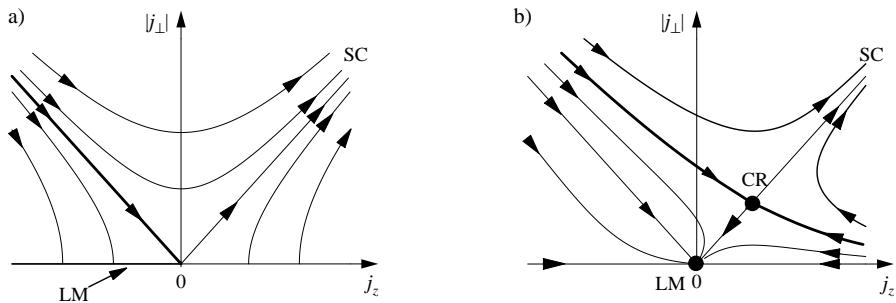}}
\caption
{
Renormalization group flow for the anisotropic Kondo model.
The thick lines denote the separatrices corresponding to the phase transitions
between local-moment (LM) and strong-coupling (SC) behavior.
a) Metallic case, i.e., for a finite bath density of states $\rho_0$ at the Fermi level.
Here the transition is of Kosterlitz-Thouless type.
b) Pseudogap case $\rho(\epsilon) = \rho_0 |\epsilon/D|^r$ ($r>0$), showing
a continuous transition with a critical fixed point CR.
}
\label{fig:kdflow}
\end{figure}

For ferromagnetic $J_K<0$, the physics is different, and no screening
occurs. Instead, in the low-temperature limit the impurity spin is
effectively decoupled from the conduction band, and contributes
a residual entropy of $S_0=k_B \ln 2$.
The two behaviors can be understood in a weak-coupling renormalization
group approach.
If we allow for spin-anisotropic Kondo interaction, and define dimensionless
couplings $j_\perp = \rho_0 J_{K\perp}$ and $j_\parallel = \rho_0 J_{K\parallel}$,
the perturbative RG flow for small couplings takes the form \cite{hewson}
\begin{equation}
\beta(j_\perp)     = j_\perp j_\parallel ~,~~
\beta(j_\parallel) = j_\perp^2~,
\end{equation}
sketched in Fig.~\ref{fig:kdflow}a.
There is a line of fixed points at $j_\perp = 0$, $j_\parallel\leq 0$,
and initial couplings with $|J_\perp| \leq -J_\parallel$ flow towards this
line, whereas for all other initial values the system flows to strong
coupling.
The line $|j_\perp|=-j_\parallel$ represents a line of transitions of the
Kosterlitz-Thouless type.
Such a transition also occurs in the classical $d\!=\!2$ XY model (Sec.~\ref{sec:KT});
it turns out that the RG equations for the Kondo problem are exactly identical
to the ones for the XY model.
In particular, no {\em critical} fixed point exists,
and power law critical behavior is absent.

This well-established picture of the Kondo effect has to be revised
for systems with vanishing conduction band DOS at the Fermi
energy.
This is the case in hard-gap systems
where $\rho(\epsilon)=0$ for energies smaller
than the gap energy, $|\epsilon|<\Delta$ \cite{shiba,hardgap},
and in so-called pseudogap systems \cite{withoff,bulla,GBI,MVLF}
with a power-law DOS $\rho(\epsilon) = \rho_0 |\epsilon/D|^r$ ($r>0$).
The former situation is realized, e.g., in semiconductors and $s$-wave superconductors,
whereas the latter arises in semimetals and in systems with long-range order
where the order parameter has nodes at the Fermi surface, e.g.,
$p$- and $d$-wave superconductors ($r=2$ and 1).

In systems with a hard gap in the conduction electron DOS,
screening is impossible at small Kondo couplings $J_K$ even
at lowest temperatures, which is easily understood by the
absence of low-energy fermions (Fig.~\ref{fig:kondovis}b).
Interestingly, in the particle-hole symmetric case screening
does not occur even for large $J_K$,
whereas in the absence of particle-hole symmetry an increase of
the Kondo coupling eventually leads to a first-order boundary
transition into a phase with Kondo screening --
this transition is characterized by a simple level crossing between
the doublet and singlet impurity states \cite{shiba,hardgap}.

\begin{figure}[!t]
\epsfxsize=6.5cm
\epsffile{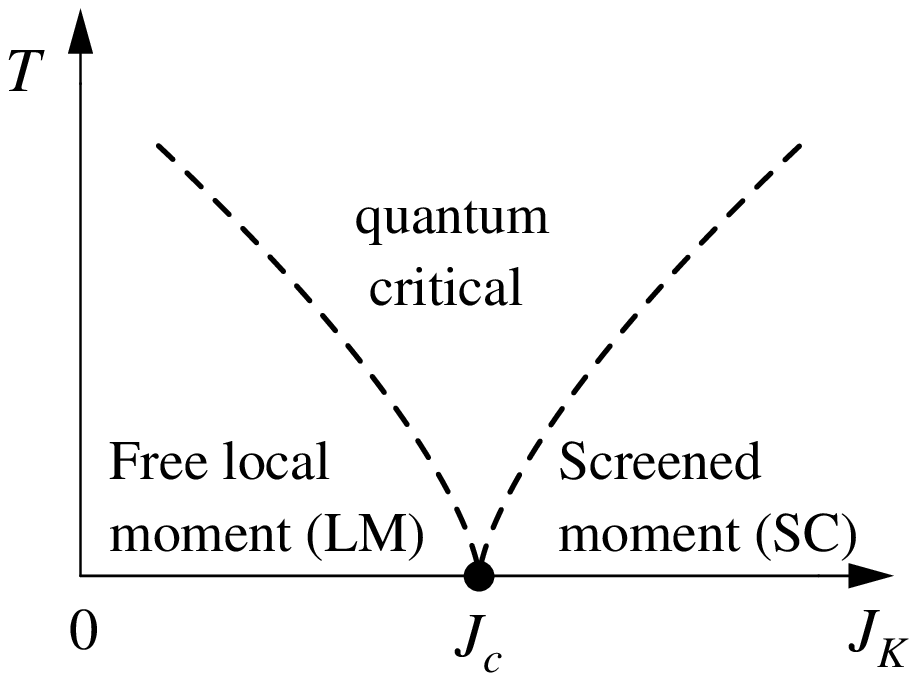}
\vspace*{-4.3cm}
\hspace*{9cm}
\epsfxsize=6cm
\epsffile{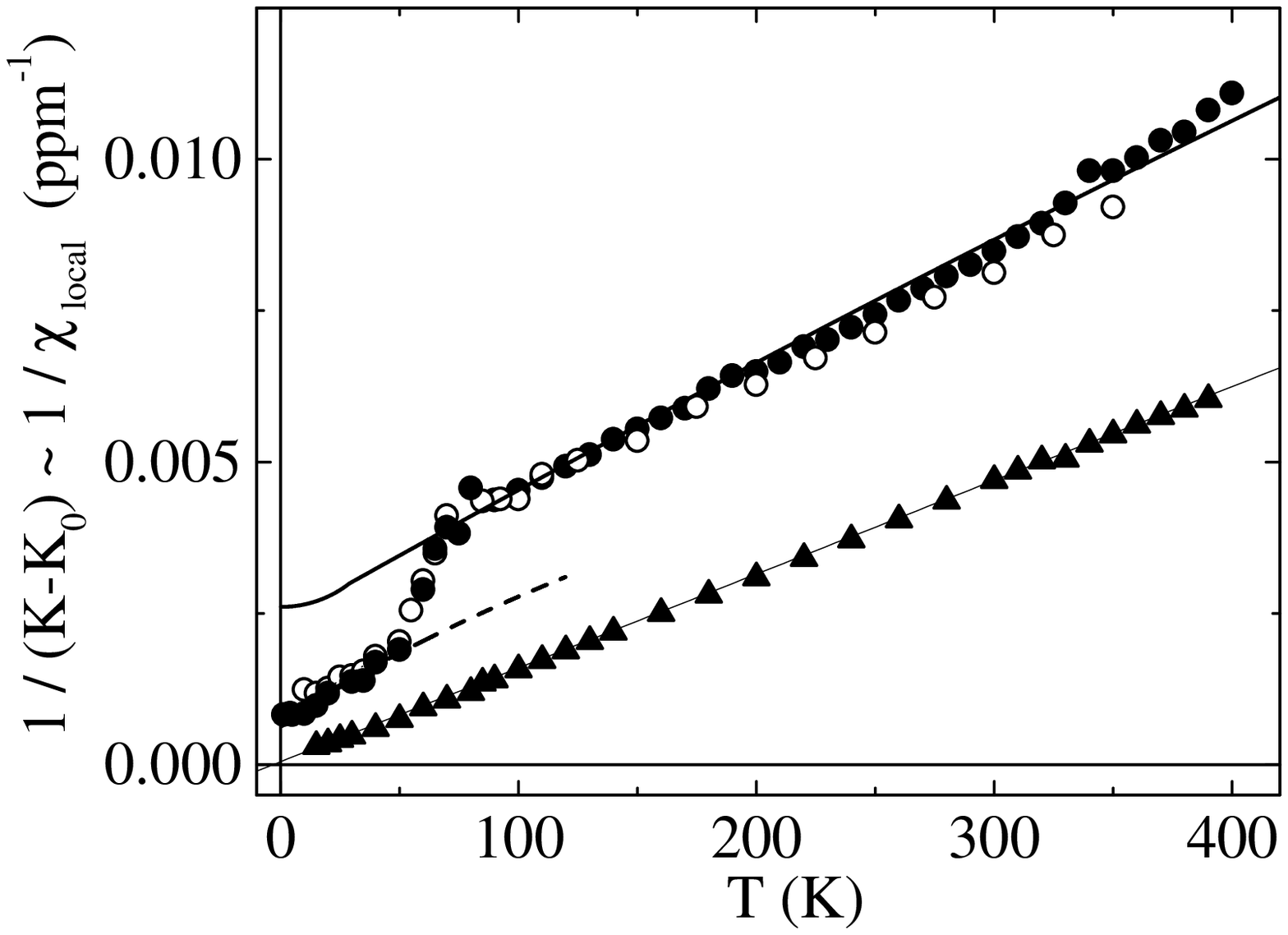}
\caption{
Left:
Phase diagram of the pseudogap Kondo model.
At $T=0$ the moment is screened at large $J_K$ (SC), but unscreened at small $J_K$ (LM).
The two phases are separated by a continuous boundary transition at $J_K=J_c$;
the quantum critical region is characterized by power laws, e.g., in the local
susceptibility and the conduction electron T matrix \cite{MVRB,PGFBK}.
The crossover line between the quantum critical and the screened regime can
be loosely associated to the Kondo temperature.
Right:
Temperature dependence of the inverse local susceptibility of dilute Li ions in
YBa$_2$Cu$_3$O$_{6+x}$ (after Ref.~\cite{bobroff}).
The circles (triangles) are data from optimally doped (underdoped) samples;
the lines are Curie-Weiss fits.
At optimal doping, the Weiss temperature changes significantly upon passing
through the superconducting $T_c$, indicating that Kondo screening above
$T_c$ is stronger than below $T_c$.
No change occurs in the underdoped sample, which should be attributed to
the pseudogap above $T_c$.
The data below $T_c$ show a Weiss temperature $(-\Theta)$ of 40 K whereas
$(-\Theta)$ is close to zero in the underdoped case -- this suppression of
Kondo screening likely corresponds to a boundary quantum phase transition.
}
\label{fig:pgk}
\end{figure}

The case of a power-law DOS is even more interesting:
the pseudogap Kondo model shows a {\em continuous} boundary quantum phase
transition at a critical Kondo coupling, $J_c$ \cite{withoff,bulla,GBI}.
For small DOS exponents $r$, the physics is captured by the RG equations
\begin{equation}
\beta(j_\perp)     = -r j_\perp     + j_\perp j_\parallel ~,~~
\beta(j_\parallel) = -r j_\parallel + j_\perp^2~,
\end{equation}
see Fig.~\ref{fig:kdflow}b.
There is now a critical fixed point at $j_\perp^\ast=j_\parallel^\ast=r$.
Initial values located below the separatrix flow to the local-moment
fixed point (LM) at $j=0$, describing an unscreened impurity,
whereas large $j$ flow to strong coupling (SC).
The resulting phase diagram is shown in Fig.~\ref{fig:pgk}a.
The behavior at large $r$ as well as the strong-coupling physics is not
described by this RG;
from comprehensive numerical studies \cite{bulla,GBI} based on Wilson's numerical
renormalization group (NRG) approach \cite{nrg}
it is known that the fixed-point structure changes for $r>1/2$, furthermore
particle-hole asymmetry is a relevant parameter in the strong-coupling regime
for $r>0$ (while being marginally irrelevant at $r=0$).

For small $r$, critical properties of the pseudogap Kondo model can be calculated
using the weak-coupling RG sketched above.
This can be understood as an expansion about the lower-critical ``dimension''
$r\!=\!0$, similar to the expansion in $(2\!+\!\epsilon)$ dimensions for the
non-linear sigma model (\ref{nlsm}).
Very recently, it has become clear that $r\!=\!1$ plays the role of the
upper-critical ``dimension'' in the pseudogap Kondo problem.
The universal critical theory in the vicinity of $r\!=\!1$ can be formulated
as a crossing of singlet and doublet impurity levels, coupled to low-energy
conduction electrons -- this is equivalent to an infinite-$U$ single-impurity
Anderson model with pseudogap DOS \cite{MVLF}.
For $r<1$ the coupling between impurity and conduction band is relevant under RG
transformations, and the model allows for a RG approach together with an expansion
in $(1\!-\!r)$,
in analogy to the expansion in $(4\!-\!\epsilon)$ dimensions of the $\phi^4$
model (\ref{phi4}).
Critical exponents can be calculated in a $(1\!-\!r)$ expansion, and hyperscaling
is obeyed.
Conversely, for $r>1$ perturbation theory is sufficient, and the transition
can be characterized as level crossing with perturbative corrections, where
hyperscaling is violated \cite{MVLF}.
We note that slave-boson mean-field theory \cite{withoff} reproduces a transition,
but does not yield sensible critical behavior.
Alternatively, a {\em dynamic} multichannel large-$N$ approach \cite{olivier}
can be applied to the pseudogap Kondo problem,
allowing for a complete analytic low-energy solution
both in the stable phases and at the quantum critical points,
with non-trivial critical behavior \cite{OSPG}.

Experimentally, signatures of non-trivial Kondo physics have been observed in
impurity-doped cuprate superconductors.
Nominally non-magnetic Zn impurities (replacing Cu) have been shown to induce
quasi-free magnetic moments in their vicinity.
The NMR data of Ref.~\cite{bobroff} allowed to fit the local impurity
susceptibility to a Curie-Weiss law, where the Weiss temperature $(-\Theta)$ can
be roughly identified with the Kondo temperature.
This experiment indicates a strongly doping-dependent Kondo temperature,
with $T_K$ in the superconducting state ranging from 40 K around optimal doping
to practically zero for strongly underdoped samples, see Fig.~\ref{fig:pgk}b.
This qualitative change in the Kondo screening properties is a good candidate
for a realization of the boundary quantum phase transition in the pseudogap Kondo
model -- there is no fundamental reason for this transition to coincide with
any of the possible {\em bulk} phase transitions in the cuprates.
As demonstrated in microscopic calculations \cite{MVRB,PGFBK},
the boundary transition from screened to unscreened dilute impurity moments in
superconducting cuprates is likely driven by the increase of the superconducting gap
upon underdoping and/or by increasing antiferromagnetic host spin fluctuations
which suppress fermionic Kondo screening (Sec.~\ref{sec:bfk}) .
Kondo screening of the Zn moments around optimal doping also leads to a characteristic
low-energy peak in the tunneling signal as measured by STM, for further discussions
see Refs.~\cite{MVRB,tolya}.

Kondo physics in a superconducting environment can be observed in
quantum dot systems, too.
A recent experiment \cite{basel} using a carbon nanotube dot coupled
to Nb leads shows a sharp crossover in the transport properties as a function
of $T_K/\Delta$, where $T_K$ is the normal-state Kondo temperature of the dot
and $\Delta$ the gap of the superconductor, consistent with the expectations
within a hard-gap Kondo model.

\subsection{Spin-boson and Bose-Kondo models}

Systems of quantum impurities coupled to {\em bosonic} baths are an equally interesting
model class;
they have been first introduced in the context of the description of dissipative
dynamics in quantum systems \cite{leggett}.
The simplest realization is the so-called spin-boson model, describing a spin or
two-level system coupled to a single bath of harmonic oscillators.
The Hamiltonian
\begin{equation}
H=-\frac{\Delta}{2}\sigma_{x}+
\sum_{i\alpha} \omega_{i}
     a_{i\alpha}^{\dagger} a_{i\alpha}
+ \sum_{i\alpha} \frac{\sigma_{\alpha}}{2}
    \lambda_{i}( a_{i\alpha} + a_{i\alpha}^{\dagger} )
\label{eq:sbm}
\end{equation}
is a straightforward generalization of the spin-boson model to
multiple baths:
the $a_{\alpha}$ are vector bosons and can be interpreted as spin-1
excitations of a ``magnetic'' bath.
The couplings between spin and bosonic baths are completely specified
by the bath spectral functions
\begin{equation}
    J_\alpha\left( \omega \right)=\pi \sum_{i}
\lambda_{i\alpha}^{2} \delta\left( \omega -\omega_{i} \right) \,.
\end{equation}
Of particular interest are power-law spectra
\begin{equation}
  J_\alpha(\omega) = 2\pi \gamma_\alpha \omega_c^{1-s} \omega^s\,,~ 0<\omega<\omega_c\,,\ \ \ s>-1
\label{power}
\end{equation}
where $\omega_c$ is a cutoff, and the dimensionless parameters
$\gamma_\alpha$ characterize the coupling or dissipation strength.

In the conventional spin-boson model only $\gamma_z\neq 0$ ($\gamma_x\!=\!\gamma_y\!=\!0$);
then Eq. (\ref{eq:sbm}) describes a spin, tunneling between
$|\uparrow\rangle$ and $|\downarrow\rangle$ via $\Delta$, and being damped
by the coupling to the oscillator bath.
The particular case of $s\!=\!1$ corresponds to the well-studied
ohmic spin-boson model \cite{leggett}, which shows a Kosterlitz-Thouless
quantum transition, separating a localized phase at $\gamma \geq \gamma_c$
from a delocalized phase at $\gamma<\gamma_c$ \cite{leggett}.
In the localized regime, the tunnel splitting between the two
levels renormalizes to zero, i.e., the system gets trapped in one
of the states $|\uparrow\rangle$ or $|\downarrow\rangle$,
whereas the tunnel splitting stays finite in the delocalized
phase.
In the limit $\Delta \ll \omega_c$ the transition occurs at $\gamma_c=1$.

Re-newed interest in spin-boson models arises in the field of quantum computation,
for modelling the coupling of qubits to a noisy environment; here also
the case of a subohmic bath appears physically relevant.
It has recently been established, using both perturbative RG for small $(1\!-\!s)$
and NRG, that the subohmic spin-boson model shows a {\em continuous} quantum transition
between a localized and a delocalized phase for all bath exponents $0<s<1$ \cite{btv}.
For fixed $s$ there exists a single critical RG fixed point with $s$-dependent exponents;
as function of $s$ these fixed points form a line which terminates in the Kosterlitz-Thouless
transition point at $s\!=\!1$.
Interestingly, this behavior is somewhat similar to the one of the particle-hole symmetric
pseudogap Kondo model \cite{bulla} described in Sec.~\ref{sec:kondo},
however, for $s<1$ and $r>0$ the transitions in the two models are in
different universality classes.

Models of the form (\ref{eq:sbm}) with a {\em vector} bath
describe impurity moments embedded into quantum magnets, and are
sometimes called Bose-Kondo models.
The bosonic baths represent the host spin excitations:
in a quantum paramagnet their spectrum is gapped, with the gap approaching
zero at a zero-temperature magnetic ordering transition.
(In dimensions $d\!<\!3$ the bath Hamiltonian has to be supplemented by bosonic
self-interactions, as those are strongly relevant, see Sec.~\ref{sec:RG}.)
At a bulk quantum critical point of an antiferromagnet in $d=2\!+\!s$ space dimensions
the bath spectra are given by Eq.~(\ref{power}).
For $1<d<3$ the impurity model has remarkable properties \cite{vbs}:
the coupling between spin and bath is relevant in the RG sense and flows to an
infrared-stable intermediate-coupling fixed point.
The impurity spin is then characterized by power-law auto-correlations and
various universal properties, e.g., a Curie response of a fractional effective
spin.
For a finite impurity concentration,
this universal interaction between impurity moments and host spin fluctuations
leads to universal impurity-induced damping of spin fluctuations
in cuprate superconductors \cite{vbs} -- note that at the relevant spin fluctuation energies
(e.g. 40 meV in YBa$_2$Cu$_3$O$_7$) possible Kondo screening of the moments
can be safely neglected, as experimentally $T_K$ is 40 K or lower \cite{bobroff}.


\subsection{Kondo effect and spin fluctuations}
\label{sec:bfk}

In particular in the context of strongly correlated electron systems,
which often feature Fermi-liquid quasiparticles and strong spin fluctuations
at the same time, the question of the interplay between fermionic
and bosonic ``Kondo'' physics arises.
This naturally leads to so-called Fermi-Bose Kondo models where
an impurity spin is coupled to {\em both} a fermionic and a bosonic bath.
The renormalization group analysis shows that the two bath couplings {\em compete},
i.e., fermionic Kondo screening is suppressed by strong host spin
fluctuations \cite{edmft,PGFBK}.

Recently it has been proposed \cite{PGFBK} that this interplay actually plays a role
in high-$T_c$ cuprates, where NMR experiments indicate that dilute impurity
moments are screened at optimal doping, but $T_K$ is essentially suppressed to
zero in underdoped compounds \cite{bobroff}, see Sec.~\ref{sec:kondo}.
A Fermi-Bose Kondo model, taking into account both the pseudogap density of states
of the Bogoliubov quasiparticles and the strong antiferromagnetic fluctuations,
provides a natural explanation: spin fluctuations increase with underdoping, thus
strongly reducing $T_K$ due to the vicinity to the boundary transition which
exists even in the absence of a bosonic bath \cite{PGFBK}.

Interestingly, similar quantum impurity models also appear in variants of
dynamical mean-field theories for lattice systems \cite{dmft}.
Motivated by neutron scattering experiments \cite{schroder} on the
heavy-fermion compound $\rm CeCu_{6-x} Au_x$
(see also Sec.~\ref{sec:metals}),
which indicate momentum-independent critical dynamics at
an antiferromagnetic ordering transition,
a self-consistent version of the Fermi-Bose Kondo model has been
proposed to describe such critical behavior within an extended
dynamical mean-field approach (EDMFT) \cite{edmft}.
In this scenario, the critical point of the lattice model is
mapped onto a particular critical point of the impurity model.


\subsection{Multi-channel and multi-impurity models}

Kondo screening is strongly modified if
two or more {\em fermionic} screening channels compete.
Nozi\`eres and Blandin \cite{Noz80} proposed a
two-channel generalization of the Kondo model, which shows
overscreening associated with an intermediate-coupling fixed point
and non-Fermi liquid behaviour
in various thermodynamic and transport properties.
In general, such behavior occurs for any number of channels $K>1$
coupled to a spin $\frac{1}{2}$, and does not require fine-tuning of
the Kondo coupling.
However, it is unstable w.r.t. a channel asymmetry;
thus a two-channel fixed point can be understood as critical point between two
(equivalent) stable single-channel fixed points.

Many of the low-energy properties of the two-channel and related Kondo models
have been studied using conformal field theory techniques \cite{olivier,AL}.
Interestingly, the multi-channel Kondo fixed point is perturbatively accessible
in the limit of large channel number ($K\gg 1$) \cite{olivier,Noz80}.
Experimental realizations have been discussed in the context
of rare-earth compounds \cite{CZ}; furthermore proposals based
on quantum-dot devices have been put forward.

Models of two or more impurities offer a new ingredient,
namely the exchange interaction, $I$, between the different
impurity spins; it can arise both from direct exchange and
from the Ruderman-Kittel-Kasuya-Yosida (RKKY) interaction mediated
by the conduction electrons.
This inter-impurity interaction competes with Kondo screening of the
individual impurities; in Kondo {\em lattice} models it can lead to a
magnetic ordering transition.
The simplest model of two spin-$\frac{1}{2}$ impurities has been
thoroughly studied:
here, a ground state singlet can be realized
either by individual Kondo screening (if $I<T_K$)
or by formation of an inter-impurity singlet (if $I>T_K$).
It has been shown that these two parameter regimes are continuously
connected (without any phase transition) as $I$ is varied
in the generic situation without particle-hole symmetry.
Notably, in the particle-hole symmetric case one finds a transition
associated with an unstable non-Fermi liquid fixed
point \cite{2impnrg,2impcft}.

Quantum phase transitions {\em generically} occur in impurity models
showing phases with {\em different} ground state spin.
For two spin-$\frac{1}{2}$ impurities, this can be realized by coupling to a
{\em single} conduction band channel only.
In this case, a Kosterlitz-Thouless-type transition between
a singlet and a doublet state occurs, associated with a secondary
exponentially small energy scale in the Kondo regime \cite{VBH}.
The physics becomes even richer if multi-channel physics
is combined with multi-impurity physics -- here, a variety
of fixed points including such with local non-Fermi liquid
behavior can be realized.

Experimentally, quantum dots provide an ideal laboratory
to study systems of two or more ``impurities'' -- note
that the local impurity states can arise either from
spin or from charge degrees of freedom on each quantum dot.
In particular, a number of experiments have been performed
on coupled quantum dot systems which can be directly mapped onto
models of two Kondo or Anderson impurities, and some
indications for phase transitions have been found \cite{wiel}.
In addition, experimental realizations of two-impurity models
using magnetic adatoms on metallic surfaces appear possible.


\section{Conclusions and outlook}
\label{sec:sum}

This review has discussed aspects of zero temperature phase transitions
in quantum systems. The introductory sections have highlighted
the general properties of these quantum transitions,
pointing out similarities and differences between classical thermal
and quantum phase transitions.
Subsequently, we have described a number of specific
bulk quantum phase transitions in condensed matter systems,
emphasizing the important role of low-energy fermionic excitations
in determining the critical behavior.
In particular, we have touched upon spin and charge ordering transitions,
secondary pairing transitions in superconductors,
as well as metal--insulator and superconductor--insulator transitions,
and made contact to recent experiments in correlated electronic
systems.
Finally, we have discussed quantum transitions in impurity
systems. Those transitions are a particular realization of
boundary quantum phase transitions, where only a part of the system becomes
critical.

Let us recapitulate a few important aspects:
The theoretical description of a particular phase transition usually starts out
with the identification of the relevant variables, the most important one being the
order parameter.
Analytical studies then often proceed following the Landau-Ginzburg-Wilson approach:
all degrees of freedom other than the order parameter fluctuations are integrated
out, resulting in an effective theory in terms of the order parameter only;
such a theory can often also be guessed using general symmetry arguments.
However, in itinerant electron systems the existence of gapless particle-hole
continua potentially leads to singularities in the order parameter theory,
rendering it non-local.
A more promising but technically difficult approach consists of {\em not}
integrating out all degrees of freedom other than the order parameter,
i.e., treating all soft modes in the system on the same footing.
This has been followed for the ferromagnetic transition of both
clean and dirty Fermi liquids \cite{sessions}, and in the somewhat
simpler situation of secondary pairing transitions in $d$-wave
superconductors \cite{vzs}.
A second caveat of analytical approaches is that they have to employ
perturbative methods in analyzing the field theories; those methods can
fail in strongly coupled or strongly disordered systems.
Alternatively, high-accuracy numerical simulations either of simple microscopic
models or of order parameter field theories can be used to access quantum
critical behavior.

Conceptually, quantum phase transitions open a field of fascinating physics,
being connected to the peculiar properties of the quantum critical ground
state.
The absence of quasiparticle excitations can lead to a variety of unsual
finite-temperature phenomena in the quantum critical region, such as unconventional
power laws and non-Fermi liquid behavior, which mask the properties of the stable
phases.
Recent years have seen an outburst of experimental activities studying
quantum criticality in systems as diverse as high-temperature
superconductors, quantum Hall systems, heavy-fermion materials, quantum
magnets, and atomic gases.
For explaining the plethora of experimental findings progress on the theory side
is required, in particular in understanding dynamical and transport properties
near quantum criticality, together with the influence of disorder.

Quantum criticality can provide new perspectives in the study of correlated
systems, where intermediate-coupling phenomena are hardly accessible by standard
weak- or strong-coupling perturbative approaches.
A promising route starts by identifying quantum critical points between stable phases,
and then uses these as vantage points for exploring the whole phase diagram by expanding
in the deviation from criticality.
It is clear that we have just scratched the surface of much exciting
progress to come.


\ack

The author is indebted to S. Sachdev for many fruitful collaborations
and countless discussions which contributed enormously to the writing of
this article.
It is pleasure to thank R. Bulla, C. Buragohain, E. Demler, L. Fritz, M. Garst, A. Georges,
S. Kehrein, M. Kir\'can, A. Polkovnikov, T. Pruschke, A. Rosch, T. Senthil, Q. Si,
N.-H. Tong, T. Vojta, D. Vollhardt, P. W\"olfle, and Y. Zhang
for illuminating conversations and collaborations
over the last years.
This work was supported by the Deutsche Forschungsgemeinschaft through
Vo794/1-1, SFB 484, and the Center for Functional Nanostructures Karlsruhe.



\end{document}